\documentclass[preprint,aps,12pt,showpacs,nofootinbib,tightenlines]{revtex4}
\usepackage{mathrsfs}
\usepackage{amsmath}
\usepackage{amssymb}
\usepackage{epsfig}
\usepackage{epstopdf}
\usepackage{graphicx}
\usepackage{subfigure}
\usepackage{booktabs}
\usepackage{float}
\usepackage{amsmath}
\usepackage{color}
\usepackage{multirow}
\usepackage{bm}
\usepackage{comment} 
\usepackage{multirow}

\def\be{\begin{eqnarray}}
\def\en{\end{eqnarray}}
\def\non{\nonumber\\}

\begin{document}
\title{The $\Upsilon(nS) \to B_{(c)}$ transition form factors and their applications to semileptonic and nonleptonic weak decays}
\author{ You-Ya Yang$^1$ Zhi-Qing Zhang$^2$ \footnote{Corresponding author. zhangzhiqing@haut.edu.cn}
Zhi-Jie Sun$^3$
 } 
\affiliation{\it \small $^1$Physics Department, College of Physics and Optoelectronic Engineering, Jinan University, Guangzhou 510632, China\\
\it \small $^2$ School of Physics and advanced energy, Henna University Technology, Zhengzhou, Henan 450001, China\\
\it \small $^3$ Bingtuan Xingxin Vocational and Technical College, Tiemenguan, Xinjiang, 841007, China} 
\date{\today}
\begin{abstract}
The semileptonic and nonleptonic decays of the $\Upsilon(nS)$ with $n=1,2,3,4$ are investigated within the covariant light-front quark model (CLFQM). Using the form factors of the transitions $\Upsilon(nS) \to B_{(c)}$ obtained from the CLFQM,  we calculate the branching ratios of the decays $\Upsilon(nS)\to B_{(c)}\ell\nu_\ell$ and $\Upsilon(nS)\to B_{(c)}M$ with $\ell=e,\mu,\tau$ and $M$ referring to $\pi(\rho),K^{(*)},D^{(*)},D^{(*)}_s$. One can find that the
branching ratios of the decays $\Upsilon(3S)\to B_c\ell\nu_\ell$ are the largest among those of considered semileptonic decays and can amount to $10^{-9}$; As to the nonleptonic decays, $\Upsilon(3S)\to B_c\rho$ and $\Upsilon(3S)\to B_cD^{(*)}_s$ have the largest branching ratios, which reach up to $10^{-10}$.  Given the identification and detection efficiency of final states, searching for these weak decay modes should be fairly challenging in future experiments.
The forward-backward asymmetry $A_{FB}$ and the longitudinal polarization fraction $f_L$ are also calculated for those semileptonic decays.
\end{abstract}

 \vspace{1cm}

\maketitle

\section{Introduction}\label{intro}
The  bottomonia $\Upsilon(1S)$, $\Upsilon(2S)$, $\Upsilon(3S)$ and $\Upsilon(4S)$ belong to the Upsilon-family, which are bound states made of $b\bar b$ with the same quantum numbers $I^G=0^-$ and $J^{PC}=1^{--}$. They have both common and different features. For $\Upsilon(1S)$, $\Upsilon(2S)$ and $\Upsilon(3S)$, their masses are near 10 GeV, and the decay widths are very narrow, only approximately $20\sim 50$ keV, while $\Upsilon(4S)$ is very different and its width reaches up to $20.5\pm2.5$ MeV. The decay final states are predominantly hadronic, amounting to approximately $95\%$, while leptonic final states account for only a few percent or even less. Especially, for the $\Upsilon(4S)$, whose mass is almost at the open bottom threshold, it decays in almost $100\%$ of cases to a $B\bar B$ pair. Their strong decays into two light hadrons via $b\bar b$ annihilation into at least three gluons are responsible for their
narrow widths. Since they were identified for the first time by the E288 Collaboration at Fermilab in 1977 \cite{herb,innes}, $\Upsilon(nS)$ have  attracted much attention from experimental and theoretical physics, which focused on their decays mediated by strong and electromagnetic interactions. The weak decays of $\Upsilon(nS)$ usually have very small branching ratios and have not been well determined in experiments up to now. The branching ratios of $\Upsilon(nS)$ inclusive weak decays can be estimated via a single quark decay of either the constituent quark or anti-quark relying on a simple spectator picture \cite{sanc}, about $2\hbar/(\tau_B\Gamma_\Upsilon)\sim\textit{O}(10^{-8})$ (even smaller for $\Upsilon(4S)$), where $\tau_B$ and $\Gamma_\Upsilon$ are the B meson lifetime and the $\Upsilon$ meson total width. For most decay channels with hadronic final states listed in the present PDG, only upper limits on the branching ratios are provided due to limited experimental data.

However, the weak decays of $\Upsilon(nS)$ mesons, although rare, offer a unique window into the underlying dynamics of heavy quarkonium weak decays. As we know, both $b$ and $\bar b$ quarks in $\Upsilon(nS)$ mesons can decay individually via the weak interaction. Furthermore, the $b$ quark weak decays serve as an ideal laboratory for testing the Standard Model and probing new physics. So the $\Upsilon(nS)$ weak decays permit one to crosscheck the parameters obtained from the $B$ meson decays. The hierarchical relationships 
of the Cabibbo-Kobayashi-Maskawa (CKM) matrix elements should also manifest in $\Upsilon(nS)$ decays, that is, the decays mediated by the $b\to c$ transition should have much larger branching ratios compared to those of the decays through the $b\to u$ transition. Currently, some phenomenological approaches have been extensively developed to deal with heavy quark weak decays, such as the perturbative QCD (PQCD) approach \cite{Yang1, Sun:2015zoa, Yang:2015iia}, Bethe-Salpeter (BS) equation \cite{T. Wang}, Covariant Confined Quark Model (CCQM) \cite{tran} and non-relativistic quantum chromodynamics (NRQCD) \cite{Sun:2015faa,Sun:2015nya,Chang:2016enr}. Except for these researches under the SM, some NP scenarios are also investigated in semileptonic weak decays $\Upsilon(nS)\to B_c \ell\nu_\ell$ \cite{Sheng}. On the experimental side, the upgraded high-luminosity LHC (HL-LHC) will produce about $10^{15}$ $b\bar b$ pairs at a center-of-mass energy of 14 TeV for a total integrated luminosity of 3000 $fb^{-1}$ \cite{Cerri}. Belle-II will have collected about $10^{10}$ $ b\bar b$ pairs by 2027 \cite{ekou}. A large amount of $\Upsilon(nS)$ data samples will offer a realistic possibility to measure some $\Upsilon(nS)$ weak decays in the near future. It is now imperative to comprehensively account for both non-leptonic and semi-leptonic decays of $\Upsilon(nS)$ within the SM framework. In this work, we will employ the CLFQM to evaluate the $\Upsilon(nS)\to B_{(c)}$ transition form factors, then research into the semi-leptonic decays $\Upsilon(nS)\to B_{(c)}\ell\nu_\ell$ and non-leptonic decays $\Upsilon(nS)\to B_{(c)}M$. 

The arrangement of this paper is as follows: In Section II, the formalism of the CLFQM, the evaluation of the $\Upsilon(nS)\to B_{(c)}$ transition form factors, the hadronic matrix elements and the helicity amplitudes for relevant decays are listed. In addition to the numerical results for the  $\Upsilon(nS)\to B_{(c)}$ transition form factors, the values for the branching ratios, the forward-backward asymmetries $A_{FB}$ and the longitudinal polarization fractions $f_L$ for the corresponding decays are presented in Section III. Detailed comparisons with other theoretical values and relevant discussions are also included. The summary is presented in Section IV. 

\section{Formalism}\label{form1}
\subsection{The form factors}
The Bauer-Stech-Wirbel (BSW) form factors for the $\Upsilon(nS)\to B$ transitions used in the fowllowing helicity ampltiudes are defined as follows,
\be
\left\langle B\left(P^{\prime \prime}\right)\left|V_{\mu}-A_{\mu}\right| \Upsilon\left(P^{\prime}, \epsilon_\Upsilon\right)\right\rangle
&=&-\epsilon_{\mu \nu \alpha \beta} \epsilon_{\Upsilon}^{\nu} q^{\alpha} P^{\beta} \frac{V\left(q^{2}\right)}{m_{\Upsilon}+m_{B}}+i \frac{2 m_{\Upsilon} \epsilon_{\Upsilon} \cdot q}{q^{2}} q_{\mu} A_{0}\left(q^{2}\right) \non
&&+i \epsilon_{\Upsilon \mu}\left(m_{\Upsilon}+m_{B}\right) A_{1}\left(q^{2}\right)+i \frac{\epsilon_{\Upsilon} \cdot q}{m_{ \Upsilon}+m_{B}}P_{\mu} A_{2}\left(q^{2}\right) \non
&&-i \frac{2 m_{\Upsilon} \epsilon_{\Upsilon} \cdot q}{q^{2}} q_{\mu} A_{3}\left(q^{2}\right),
\en
where $P=P'+P'', q=P'-P''$ and the convention $\epsilon_{0123}=1$ is adopted. $A_{3}(q^2)$ is related to $A_{1}(q^2)$ and $A_{2}(q^2)$ as
\be
A_{3}(q^2)=\frac{\left(m_{\Upsilon}+m_{B}\right)}{2 m_{\Upsilon}} A_{1}(q^2)+\frac{\left(m_{ \Upsilon}-m_{B}\right)}{2 m_{\Upsilon}} A_{2}(q^2).
\en
While the Isgur-Scora-Grinstein-Wise (ISGW) defination for these transitions is given as
	\be
\langle B (P^{\prime\prime})|(V-A)_\mu|\Upsilon(P^\prime,\epsilon_\Upsilon)\rangle &=&g(q^2)\varepsilon_{\mu\nu\alpha\beta}\epsilon^{\nu}_\Upsilon P^\alpha q^\beta-i\left\{f(q^2)\epsilon_{\Upsilon\mu}\right.\\
&&\left.+\epsilon_{\Upsilon}\cdot P\left[P_\mu a_+(q^2)+q_\mu a_-(q^2)\right]\right\}.
\en

\begin{figure}[htbp]
\centering \subfigure{
\begin{minipage}{5cm}
\centering
\includegraphics[width=5cm]{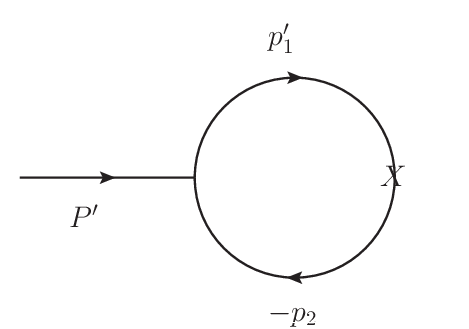}
\end{minipage}}
\subfigure{
\begin{minipage}{6cm}
\centering
\includegraphics[width=6cm]{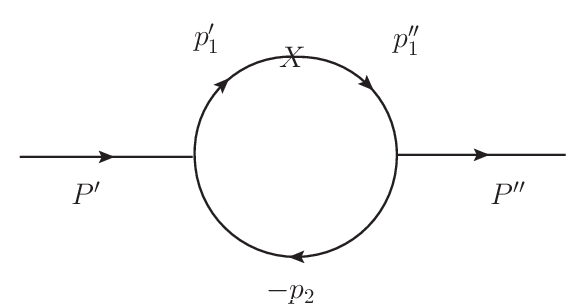}
\end{minipage}}
\caption{Feynman diagrams for bottomonium decay (left) and transition
(right) amplitudes, where $P^{\prime(\prime\prime)}$ is the
incoming (outgoing) meson momentum, $p^{\prime(\prime\prime)}_1$
is the quark momentum, $p_2$ is the anti-quark momentum and X
denotes the vector or axial-vector transition vertex.}
\label{feyn}
\end{figure}

The results of the lowest order form factors could be obtained by calculating the Feynman diagrams shown in Figure \ref{feyn}. In the covariant quark model, the treatment of transition form factors is relatively covariant throughout the calculation process, where the light-front coordinates of a momentum $p$ are used $p=(p^-,p^+,p_\perp)$ with
$p^\pm=p^0\pm p_z, p^2=p^+p^--p^2_\perp$.
The incoming (outgoing) meson has the mass $M^\prime(M^{\prime\prime})$
with the momentum $P^\prime=p_1^\prime+p_2 (P^{\prime\prime}=p_1^{\prime\prime}+p_2)$, where $p_{1}^{\prime(\prime\prime)} $
and $p_{2}$ are the momenta of the quark and anti-quark
inside the incoming (outgoing) meson with the mass $m_{1}^{\prime(\prime\prime)}$and $m_{2}$, respectively. Here we use the same notations
as those in Refs. \cite{jaus,Y. Cheng} and $M^\prime$ refers to the $\Upsilon(nS)$ masses.
These momenta can be expressed in terms of the internal variables $(x_{i},p{'}_{\perp})$ as
\be
p_{1,2}^{\prime+}=x_{1,2} P^{\prime+}, \quad p_{1,2 \perp}^{\prime}=x_{1,2} P_{\perp}^{\prime} \pm p_{\perp}^{\prime},
\en
with $x_{1}+x_{2}=1$  . Using these internal variables,
we can define some quantities for the incoming meson which will be used in the following calculations
\be
M_{0}^{\prime 2} &=&\left(e_{1}^{\prime}+e_{2}\right)^{2}=\frac{p_{\perp}^{\prime 2}+m_{1}^{\prime 2}}{x_{1}}
+\frac{p_{\perp}^{2}+m_{2}^{2}}{x_{2}}, \quad \widetilde{M}_{0}^{\prime}=\sqrt{M_{0}^{\prime 2}-\left(m_{1}^{\prime}-m_{2}\right)^{2}},\non
e_{i}^{(\prime)} &=&\sqrt{m_{i}^{(\prime) 2}+p_{\perp}^{\prime 2}+p_{z}^{\prime 2}}, \quad \quad p_{z}^{\prime}
=\frac{x_{2} M_{0}^{\prime}}{2}-\frac{m_{2}^{2}+p_{\perp}^{\prime 2}}{2 x_{2} M_{0}^{\prime}},
\en
where $M'_0$ is the kinetic invariant mass of the incoming meson and can be expressed as the energies of the quark and the anti-quark
$e^{(\prime)}_i$. It is similar to the case of the outgoing meson. In order to calculate the amplitudes of the transition form factors, we need the following Feynman rules for the meson-quark-antiquark vertex $(i\Gamma^{'} _{P,V})$:
\be
i\Gamma^{'} _{P}&=&H^{'}_{P}\gamma_{5},\\
i\Gamma^{'} _{V}&=&i H_{V}^{\prime}\left[\gamma_{\mu}-\frac{1}{W_{V}^{\prime}}\left(p_{1}^{\prime}-p_{2}\right)_{\mu}\right].
\en
For the general $\Upsilon(nS)\to B$ transition, the decay amplitude for the lowest order is
\be
\mathcal{B}_{\mu}^{\Upsilon B }=-i^{3} \frac{N_{c}}{(2 \pi)^{4}} \int d^{4} p_{1}^{\prime} \frac{H_{\Upsilon}^{\prime}H_{ B }^{\prime\prime}}
{N_{1}^{\prime} N_{1}^{\prime \prime} N_{2}} S_{\mu\nu}^{\Upsilon  B}\varepsilon^{*\nu}, \label{etacD}
\en
where $N_{1}^{\prime(\prime \prime)}=p_{1}^{\prime(\prime \prime) 2}-m_{1}^{\prime (\prime\prime) 2}, N_{2}=p_{2}^{2}-m_{2}^{2} $ arise
from the quark propagators, and
the trace $S_{\mu\nu}^{\Upsilon  B}$ can be obtained directly by using the Lorentz contraction,
\be
S_{\mu \nu}^{\Upsilon B}&=&\left(S_{V}^{\Upsilon B}-S_{A}^{\Upsilon B}\right)_{\mu \nu}\non
&=&\operatorname{Tr}\left[\left(\gamma_{\nu}-\frac{1}{W_{V}^{\prime}}\left(p_{1}^{\prime}-p_{2}\right)_{\nu}\right)\left(p_{1}^{\prime \prime}
+m_{1}^{\prime \prime}\right)\left(\gamma_{\mu}-\gamma_{\mu} \gamma_{5}\right)\left(\not p_{1}^{\prime}+m_{1}^{\prime}\right) \gamma_{5}\left(-\not p_{2}
+m_{2}\right)\right].
\label{sptov}
\en
The specific expressions for $S_{\mu \nu}^{\Upsilon B}$ can be found in Appendix A.
In practice, we use the light-front decomposition of the Feynman loop momentum and integrate out
the minus component through the contour method. The specific rules for such integration are displayed in Appendix A. If the covariant vertex functions are not singular when performing integration,
the transition amplitudes will
pick up the singularities in the anti-quark propagators. The integration then leads to
\be
W_{\Upsilon}^{\prime} &\rightarrow& w_{\Upsilon}^{\prime}=M^{'}_{0}+m^{'}_{1}+m_{2}, \non
\int \frac{d^{4} p_{1}^{\prime}}{N_{1}^{\prime} N_{1}^{\prime \prime} N_{2}} H_{\Upsilon} H_{B} S^{\Upsilon B}_{\mu\nu} & \rightarrow&-i \pi \int \frac{d x_{2} d^{2}
p_{\perp}^{\prime}}{x_{2} \hat{N}_{1}^{\prime} \hat{N}_{1}^{\prime \prime}} h_{\Upsilon} h_{B} \hat{S}^{\Upsilon B}_{\mu\nu},
\en
where
\be
\hat{N}_{1}^{\prime(\prime \prime)}&=&x_{1}\left(M^{\prime(\prime \prime)2}-M_{0}^{\prime(\prime \prime) 2}\right),\non
h_{\Upsilon(B)} &=&\left(M^{\prime(\prime\prime) 2}-M_{0}^{\prime(\prime\prime) 2}\right) \sqrt{\frac{x_{1} x_{2}}{N_{c}}} \frac{1}{\sqrt{2} \widetilde{M}_{0}^{\prime(\prime\prime)}} \varphi.
\label{vertex}
\en
It is noticed that $M^{\prime\prime 2}_0$ is defined as 
\be
M_{0}^{\prime\prime 2} &=&\frac{p_{\perp}^{\prime\prime 2}+m_{1}^{\prime\prime 2}}{x_{1}}
+\frac{p_{\perp}^{2}+m_{2}^{2}}{x_{2}}
\en
with $p''_\perp=p'_\perp-x_2q_\perp$, and $\varphi$ is light-front momentum distribution amplitude for S-wave mesons 
\be
\varphi &=&\varphi\left(x, p_{\perp}\right)=4\left(\frac{\pi}{\beta^{2}}\right)^{\frac{3}{4}}
\sqrt{\frac{d p_{z}}{d x}} \exp \left(-\frac{p_{z}^{2}+p_{\perp}^{ 2}}{2 \beta^{2}}\right),
\en
where $\beta$ is a phenomenological parameter and can be fixed by fitting the corresponding decay constant. As to the radially excited states $\Upsilon(2S),\Upsilon(3S)$ and $\Upsilon(4S)$, the distribution functions are given as
\be
  \varphi(2S)&=& 4(\frac{\pi}{\beta^{2}})^{3/4}\sqrt{\frac{d p_{z}}{d x}}\mathrm{exp}(-\frac{p_{z}^{2}+p_{\perp}^{2}}{2\beta^{2}})\frac{1}{\sqrt{6}}(-3+2\frac{p_z^2+p_\perp^2}{\beta^2}), \\
  \varphi(3S)&=& 4(\frac{\pi}{\beta^2})^{3/4}\sqrt{\frac{d p_{z}}{d x}}\mathrm{exp}(-\frac{p_z^2+p_\perp^2}{2\beta^2}) \frac{1}{2\sqrt{30}}(-15-20\frac{p_{z}^{2}+p_{\perp}^{2}}{\beta^{2}}+4\frac{(p_{z}^{2}+p_{\perp}^{2})^{2}}{\beta^{4}}), 
\en
\be
  \varphi(4S)&=& 4(\frac{\pi}{\beta^{2}})^{3/4}\sqrt{\frac{d p_{z}}{d x}}\mathrm{exp}(-\frac{p_{z}^{2}+p_{\perp}^{2}}{2\beta^{2}})\frac{1}{12\sqrt{35}}(- 105+210\frac{p_{z}^{2}+p_{\perp}^{2}}{\beta^{2}}\notag\\
  &&-84\frac{(p_{z}^{2}+p_{\perp}^{2})^{2}}{\beta^{4}}
+8\frac{(p_z^2+p_\perp^2)^3}{\beta^6}).
\en

Matching Eq.(\ref{etacD}) with the general $V\to P$ transition expression given by the ISGW definition and combining the integration rules listed in Appendix B, we can obtain the corresponding form factors, which are related to the BSW ones
through the following relations,
\be
V(q^2)&=&-(m_{\Upsilon}+m_{B})g(q^2),\;\;\;\;\; A_1(q^2)=-\frac{f(q^2)}{m_{\Upsilon}+m_{B}}, \label{relationv1}\\
A_2(q^2)&=&(m_{\Upsilon}+m_{B})a_+(q^2),A_3(q^2)-A_0(q^2)=\frac{q^2}{2m_{B}}a_-(q^2). \label{relationv2}
\en
The specific expressions for these form factors can be found in Appendix C.
\subsection{Helicity amplitudes and Observables}\label{decaycons}
Combining the helicity amplitudes via the form factors derived from the CLFQM, one can calculate the differential decay widths of the semileptonic decays $\Upsilon(nS)\to B\ell\nu_\ell$, which are listed as follows:
 \begin{footnotesize}
\begin{eqnarray}
 \frac{d\Gamma_L(\Upsilon\to B\ell\nu_\ell)}{dq^2}&=&(\frac{q^2-m_\ell^2}{q^2})^2\frac{ {\sqrt{\lambda(m_{\Upsilon}^2,m_{B}^2,q^2)}} G_F^2 |V_{qb}|^2} {384m_{\Upsilon}^3\pi^3q^2}
  \left\{ 3 m_\ell^2 \lambda(m_{\Upsilon}^2,m_{B}^2,q^2) A_0^2(q^2)\right.+(m_\ell^2+2q^2)\nonumber\\
 && \times\left.\left|\frac{1}{2m_{B}}  \left[
 (m_{\Upsilon}^2-m_{B}^2-q^2)(m_{\Upsilon}+m_{B})A_1(q^2)-\frac{\lambda(m_{\Upsilon}^2,m_{B}^2,q^2)}{m_{\Upsilon}+m_{B}}A_2(q^2)\right]\right|^2
 \right\},\label{eq:decaywidthlon}\;\;\;\;\;\;\\
\frac{d\Gamma_\pm(\Upsilon\to B\ell\nu_\ell)}{dq^2}&=&(\frac{q^2-m_\ell^2}{q^2})^2\frac{ {\sqrt{\lambda(m_{\Upsilon}^2,m_{B}^2,q^2)}} G_F^2 |V_{qb}|^2} {384m_{\Upsilon}^3\pi^3}
  \nonumber\\
 &&\;\;\times \left\{ (m_\ell^2+2q^2) \lambda(m_{\Upsilon}^2,m_{B}^2,q^2)\left|\frac{V(q^2)}{m_{\Upsilon}+m_{B}}\mp
 \frac{(m_{\Upsilon}+m_{B})A_1(q^2)}{\sqrt{\lambda(m_{\Upsilon}^2,m_{B}^2,q^2)}}\right|^2
 \right\},\label{eq:widthlon2}
\end{eqnarray}
\end{footnotesize}
where $G_{F}$ is the Fermi coupling constant, the CKM matrix element $V_{qb}$ with $q=u,c$,  $\lambda(m^{2}_{\Upsilon},m^{2}_{B},q^{2})=(m^{2}_{\Upsilon}+m^{2}_{B}-q^{2})^{2}-4m^{2}_{\Upsilon}m^{2}_{B}$ and $m_{\ell}$ is the mass of the lepton. 
The combined transverse and total differential decay widths are defined as
\be
\frac{d \Gamma_{T}}{d q^{2}}=\frac{d \Gamma_{+}}{d q^{2}}+\frac{d \Gamma_{-}}{d q^{2}}, \quad \frac{d \Gamma}{d q^{2}}=\frac{d \Gamma_{L}}{d q^{2}}+\frac{d \Gamma_{T}}{d q^{2}}.
\en

For the semileptonic decays $\Upsilon(nS)\to B\ell\nu_\ell$ decays, it is meaningful to define the polarization fraction due to the existence of different polarizations
\be
f_{L}=\frac{\Gamma_{L}}{\Gamma_{L}+\Gamma_{+}+\Gamma_{-}}. \label{eq:fl}
\en
As to the forward-backward asymmetry, the analytical expression is defined as \cite{Sakaki:2013bfa}
\be
A_{FB} = \frac{\int^1_0 {d\Gamma \over dcos\theta} dcos\theta - \int^0_{-1} {d\Gamma \over dcos\theta} dcos\theta}
{\int^1_{-1} {d\Gamma \over dcos\theta} dcos\theta} = \frac{\int b_\theta(q^2) dq^2}{\Gamma_{\Upsilon\to B\ell\nu_\ell}},\label{eq:AFB}
\en
where $\theta$ is the angle between the 3-momenta of the lepton $\ell$ and the initial $\Upsilon(nS)$ meson in the $\ell\nu_\ell$ rest frame. The function $b_{\theta}(q^2)$ represents the angular coefficient and is written as \cite{Sakaki:2013bfa}
\be
b_\theta(q^2) &=& {G_F^2 |V_{qb}|^2 \over 128\pi^3 m_{\Upsilon}^3} q^2 \sqrt{\lambda(q^2)}
\left( 1 - {m_\ell^2 \over q^2} \right)^2 \left[ {1 \over 2}(H_{V,+}^2-H_{V,-}^2)+ {m_\ell^2 \over q^2} ( H_{V,0}H_{V,t} ) \right],
\label{eq:btheta2}
\en
where the helicity amplitudes
\be
H_{V,\pm}\left(q^{2}\right)&=&\left(m_{\Upsilon}+{m_{B}}\right) A_{1}\left(q^{2}\right) \mp \frac{\sqrt{\lambda\left(q^{2}\right)}}{m_{\Upsilon}+m_{B}} V\left(q^{2}\right), \non
H_{V,0}\left(q^{2}\right)&=&\frac{m_{\Upsilon}+m_{B}}{2 m_{\Upsilon} \sqrt{q^{2}}}\left[-\left(m_{\Upsilon}^{2}-m_{B}^{2}-q^{2}\right) A_{1}\left(q^{2}\right)+\frac{\lambda\left(q^{2}\right) A_{2}\left(q^{2}\right)}{\left(m_{\Upsilon}+m_{B}\right)^{2}}\right],\non
H_{V,t}\left(q^{2}\right)&=&-\sqrt{\frac{\lambda\left(q^{2}\right)}{q^{2}}} A_{0}\left(q^{2}\right).
\en
Here the subscript $V$ in each helicity amplitude refers to the $\gamma_\mu(1-\gamma_5)$ current.
\subsection{Hadronic matrix elements}
In phenomenology, the effective Hamiltonian of the nonleptonic weak decays $\Upsilon(nS)\to B_{(c)}M$ can be written as \cite{Buchalla}
\be
\mathcal{H}_{\mathrm{eff}}=\frac{G_{F}}{\sqrt{2}} \sum_{q^\prime=d,s} V_{u(c) b}^{*} V_{q q^\prime}\left\{C_{1}(\mu) Q_{1}(\mu)+C_{2}(\mu) Q_{2}(\mu)\right\}+\text { h.c. },
\en
where  $V^{\ast}_{u(c)b}V_{qq^\prime}$ is the product of the CKM matrix elements with $q=u,c$ and $q^\prime=s,d$, and $C_{1,2}(\mu)$ are the Wilson coefficients.
The local tree four-quark operators $Q_{1,2}$ are defined by:
\be
Q_{1}&=&\left[\bar{q}_{ \alpha} \gamma_{\mu}\left(1-\gamma_{5}\right) b_{\alpha}\right]\left[\bar q^\prime_{\beta} \gamma^{\mu}\left(1-\gamma_{5}\right) q_{\beta}\right], \\
Q_{2}&=&\left[\bar{q}_{ \alpha} \gamma_{\mu}\left(1-\gamma_{5}\right) b_{\beta}\right]\left[\bar q^\prime_{ \alpha} \gamma^{\mu}\left(1-\gamma_{5}\right) q_{\beta}\right],
\en
where $\alpha$ and $\beta$ are color indices. 
In addition, the amplitudes for the decays $\Upsilon(nS) \rightarrow B_{(c)}P$ with $P= \pi, K, D_{(s)}$ can be expressed as:
\be
\mathcal{A}( \Upsilon \rightarrow B_{(c)} P)=\left\langle B_{(c)} P\left|\mathcal{H}_{e f f}\right| \Upsilon\right\rangle=\frac{G_{F}}{\sqrt{2}} V_{u(c)b}^{*} V_{qq^\prime} a_{1} 2 m_{\Upsilon}\left(\epsilon_{\Upsilon} \cdot q\right) f_{P} A_{0}\left(m_{P}^{2}\right).
\en
For the $ \Upsilon(nS)\rightarrow B V$ with $V=\rho,K^*$ and $D^*_{(s)}$ decays, the hadronic matrix elements can be expressed as
\be
\mathcal{A}\left(  \Upsilon \rightarrow B_{(c)} V \right)=\left\langle B_{(c)} V\left|\mathcal{H}_{\mathrm{eff}}\right|   \Upsilon\right\rangle=\frac{G_{F}}{\sqrt{2}} V_{u(c)b}^{*} V_{q q^\prime} a_{1} H_{\lambda},
\en
where $\lambda$ denotes the helicity of the vector meson, and $\mathcal{H}_{\lambda}=\left\langle V\left|J^{\mu}\right| 0\right\rangle\left\langle B_{(c)} \left|J_{\mu}\right|   \Upsilon\right\rangle$ is given as follows
\be
H_{0} &\equiv&  \left\langle V\left(\varepsilon_{V0}, p_{V}\right)\left|\bar{q}^\prime \gamma^{\mu} q\right| 0\right\rangle\left\langle B\left(p_{B}\right)\left|\bar{q} \gamma_{\mu}\left(1-\gamma_{5}\right) b\right|  \Upsilon\left(\varepsilon_{\Upsilon0}, p_{  \Upsilon}\right)\right\rangle \non
&=&\frac{i f_{V}}{2 m_{  \Upsilon}}\left[\left(m_{  \Upsilon}^{2}-m_{B}^{2}+m_{V}^{2}\right)\left(m_{  \Upsilon}+m_{B}\right) A_{1}\left(m_{V}^{2}\right)\right. \non&&
 \left.+\frac{4 m_{  \Upsilon}^{2} p_{c}^{2}}{m_{  \Upsilon}+m_{B}} A_{2}\left(m_{V}^{2}\right)\right], \\
H_{\pm} &\equiv& \left\langle V\left(\varepsilon_{V\pm}, p_{V}\right)\left|\bar{q}^\prime \gamma^{\mu} q\right| 0\right\rangle\left\langle B\left(p_{B}\right)\left|\bar{q} \gamma_{\mu}\left(1-\gamma_{5}\right) b\right|   \Upsilon\left(\varepsilon_{\Upsilon\pm}, p_{  \Upsilon}\right)\right\rangle \non
&=&i f_{V} m_{V}\left[-\left(m_{ \Upsilon}+m_{B}\right) A_{1}\left(m_{V}^{2}\right) \pm \frac{2 m_{ \Upsilon} p_{c}}{m_{ \Upsilon}+m_B} V\left(m_{V}^{2}\right)\right].
\en
\section{Numerical results and discussions} \label{numer}
\subsection{Transition Form Factors}
\begin{table}[H]
	\caption{Decay constants $f_{\Upsilon}$ obtained from the branching ratios of the leptonic decays $\Upsilon(nS)\to \ell^+\ell^-$, and the last column "$\bar f_{\Upsilon}$" refers to the weighted averages of $f_{\Upsilon}$.}
	\begin{center}
		\scalebox{0.75}{
			\begin{tabular}{|c|c|c|c|}
				\hline
				decay mode &$\mathcal{B}r$ \cite{pdg24}& $f_{\Upsilon}$ & $\bar f_{\Upsilon}$\\
				\hline
				$\Upsilon(1S)\to e^{+}e^{-}$&$(2.38 \pm 0.11)\%$&$700.4\pm32.4$ &$$    \\
				$\Upsilon(1S)\to \mu^{+}\mu^{-}$&$(2.48\pm0.05)\%$&$714.9\pm14.4$   &$716.0\pm11.9$\\
				$\Upsilon(1S)\to \tau^{+}\tau^{-}$&$(2.60\pm0.10)\%$&$ 732.0\pm28.2$   &$$  \\
				\hline
				$\Upsilon(2S)\to e^{+}e^{-}$&$(1.91 \pm 0.16)\%$&$496.9\pm41.6$    &$$ \\
				$\Upsilon(2S)\to \mu^{+}\mu^{-}$&$(1.93\pm0.17)\%$&$499.5\pm44.0$  &$500.6\pm26.3$ \\
				$\Upsilon(2S)\to \tau^{+}\tau^{-}$&$(2.00\pm0.21)\%$&$ 508.5\pm53.4$  &   \\
				\hline
				$\Upsilon(3S)\to e^{+}e^{-}$&$(2.18 \pm 0.20)\%$&$430.1\pm39.5$  &$$   \\
				$\Upsilon(3S)\to \mu^{+}\mu^{-}$&$(2.18\pm0.21)\%$&$430.1\pm41.4$  &$432.2\pm25.6$ \\
				$\Upsilon(3S)\to \tau^{+}\tau^{-}$&$(2.29\pm0.30)\%$& $440.8\pm57.7$    & \\
				\hline
				$\Upsilon(4S)\to e^{+}e^{-}$&$(1.57 \pm 0.08)\times10^{-5}$&$370.6\pm18.9$    & \\
				$\Upsilon(4S)\to \mu^{+}\mu^{-}$& &  &$370.6\pm18.9$ \\
				$\Upsilon(4S)\to \tau^{+}\tau^{-}$& &  &   \\
				\hline
			\end{tabular}
			\label{f}
		}
	\end{center}
\end{table}
The decay constants $f_{\Upsilon}$ of $\Upsilon(nS)$ can be extracted from the processes $\Gamma(\Upsilon(nS)\to \ell^{+}\ell^{-})$ with
\be
\Gamma\left(\Upsilon(nS) \rightarrow \ell^{+} \ell^{-}\right)=\frac{4 \pi}{27} \frac{\alpha^{2}}{m_{\Upsilon}} f_{\Upsilon}^{2}\sqrt{1-2 \frac{m_{\ell}^{2}}{m_{\Upsilon}^{2}}}(1+2 \frac{m_{\ell}^{2}}{m_{\Upsilon}^{2}}).
\en
Using the experimental data from PDG \cite{pdg24} as inputs, we obtain the values of each decay constant $f_{\Upsilon}$, which are listed in Table\ref{f}.

\begin{table}[H]
	\caption{The values of the input parameters \cite{Zhang:2023ypl,pdg24,damir,chiu,Wingate}. }
	\label{tab:constant}
	\begin{tabular*}{16.5cm}{@{\extracolsep{\fill}}l|cccccc}
		\hline\hline
		\textbf{Mass(\text{GeV})} &$m_{b}=4.8$
		&$m_{c}=1.4$&$m_{s}=0.37$&$m_{u,d}=0.25$&$m_{\mu}=0.106$   \\[1ex]
		&$m_{\pi}=0.140$&$m_{K}=0.494$&$m_{\rho}=0.775$&$m_{K^{\star}}=0.892$& $m_\tau=1.78$&\\[1ex]
		& $m_{\Upsilon(1S)}=9.460$& $ m_{\Upsilon(2S)}=10.023$  & $m_{\Upsilon(3S)}=10.355 $& $m_{\Upsilon(4S)}=10.579$& $m_{B_{c}}=6.27447 $ \\[1ex]
		& $ m_B=5.279$  & $m_D=1.86966 $& $m_{D^*}=2.010$& $m_{D_s}=1.96835$& $m^*_{D_s}=2.1122$\\ [1ex]
		\hline
	\end{tabular*}
	\begin{tabular*}{16.5cm}{@{\extracolsep{\fill}}l|ccccc}
		\hline
		\multirow{2}{*}{{\textbf{CKM(\text{GeV})}}}&$V_{cd}=(0.221\pm0.004)$&$V_{us}=(0.2243\pm0.0008)$\\[1ex]
		& $V_{ud}=(0.97373\pm0.00031)$&$V_{cs}=(0.975\pm0.006)$ \\[1ex]
		& $V_{ub}=(0.00382)$&$V_{cb}=(40.8\pm1.4)\times10^{-3}$ \\[1ex]
		\hline
	\end{tabular*}
	\begin{tabular*}{16.5cm}{@{\extracolsep{\fill}}l|ccccc}
		\hline
		\textbf{ decay constants(\text{GeV})} & $f_{\pi}=0.132$ &  $f_{\Upsilon(1S)}=0.716\pm0.0119$\\[1ex]
		&  $f_{K}=0.16$
		&$f_{\Upsilon(2S)}=0.5006\pm0.0263$\\[1ex]
		& $f_{D}=0.2046\pm0.005$ & $f_{\Upsilon(3S)}=0.4322\pm0.0256$\\[1ex]
		& $f_{D^*}=0.245\pm0.02^{+0.003}_{-0.002}$  &$f_{\Upsilon(4S)}=0.3706\pm0.0189$\\[1ex]
		& $f_{\rho}=0.209$ 
		& $f_{D_s}=0.2575\pm0.0046$\\[1ex]
		& $f_{K^{*}}=0.217$  
		& $f_{B_{c}}=0.489$ \\[1ex] &$f_{B}=0.19$ &$f_{D^{*}_s}=0.272\pm0.016^{+0.003}_{-0.002}$\\[1ex]
		\hline\hline
	\end{tabular*}
	\begin{tabular*}{16.5cm}{@{\extracolsep{\fill}}l|ccccc}
		\textbf{shape parameters(\text{GeV})}&$\beta^{'}_{\Upsilon(1S)}=1.289\pm0.014$&$\beta^{'}_{\Upsilon(2S)}=0.920^{+0.031}_{-0.032}$&$\beta^{'}_{\Upsilon(3S)}=0.808\pm0.032$\\[1ex]
		&$\beta^{'}_{\Upsilon(4S)}=0.687^{+0.022}_{-0.023}$& $\beta^{'}_{B_{c}}=1.058^{+0.009}_{-0.010}$&$\beta^{'}_{B}=0.555^{+0.048}_{-0.048}$\\[1ex]
		\hline\hline
	\end{tabular*}
	\begin{tabular*}{16.5cm}{@{\extracolsep{\fill}}l|ccc}
		\textbf{Full width}&$\Gamma_{\Upsilon(1S)}=(54.02\pm1.25)$ keV&$\Gamma_{\Upsilon(2S)}=(31.98\pm2.63)$ keV\\[1ex]
		$$&$\Gamma_{\Upsilon(3S)}=(20.32\pm1.85)$ keV&$\Gamma_{\Upsilon(4S)}=(20.5\pm2.5)$ MeV\\[1ex]
		\hline\hline
	\end{tabular*}
\end{table}

Obviously, the decay constants of $\Upsilon(nS)$ decrease 
with $n$. Using these decay constants, one can determine the shape parameters $\beta_\Upsilon$ shown in Table \ref{tab:constant}.
Other input parameters, including the masses of the initial and final mesons, the CKM matrix elements, the decay constants and the full widths of $\Upsilon(nS)$ are listed in Table \ref{tab:constant}. 
Based on these input parameters, the numerical results of the transition form factors can be obtained at $q^2=0$, which are listed in Tables \ref{formfactor} and  \ref{formfactor2}.

\begin{figure}[H]
	\vspace{0.32cm}
	\centering
	\subfigure[]{\includegraphics[width=0.20\textwidth]{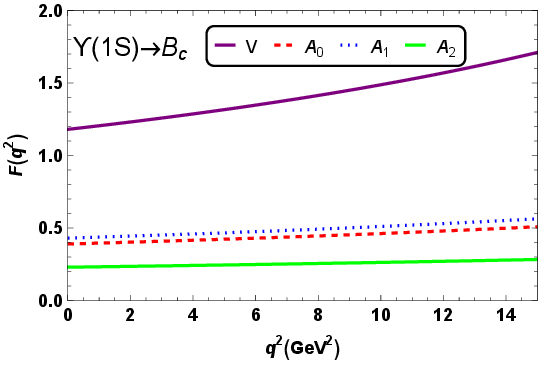}}
	\subfigure[]{\includegraphics[width=0.20\textwidth]{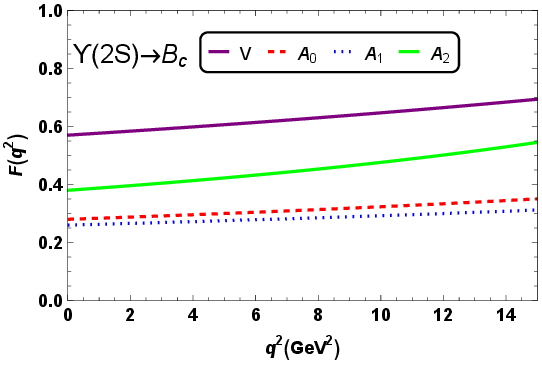}}
	\subfigure[]{\includegraphics[width=0.20\textwidth]{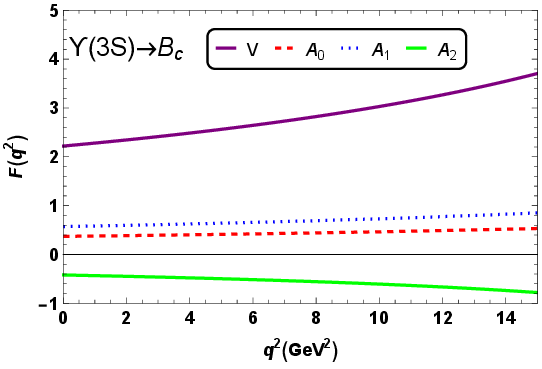}}
	\subfigure[]{\includegraphics[width=0.20\textwidth]{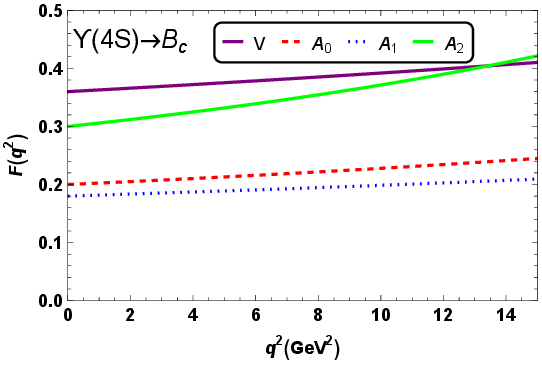}}\\
	\subfigure[]{\includegraphics[width=0.20\textwidth]{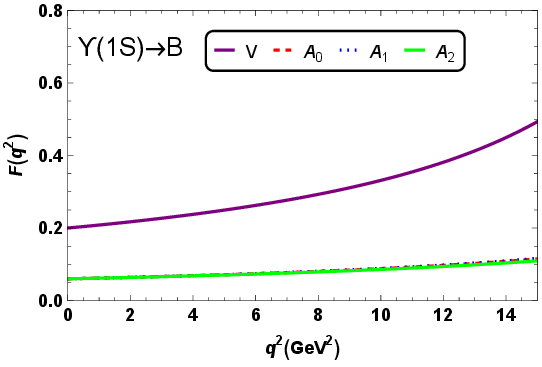}}
	\subfigure[]{\includegraphics[width=0.20\textwidth]{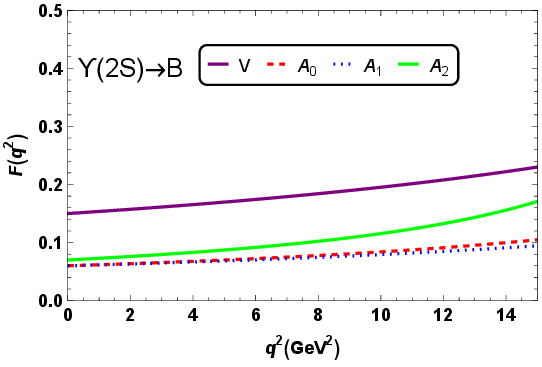}}
	\subfigure[]{\includegraphics[width=0.20\textwidth]{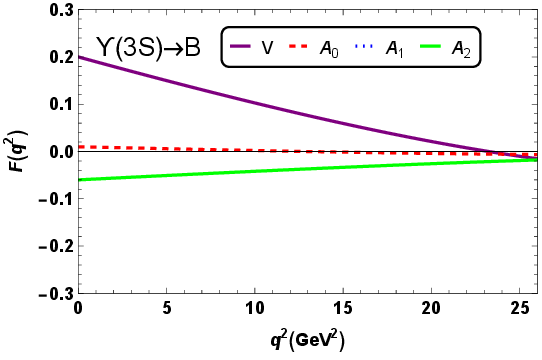}}
	\subfigure[]{\includegraphics[width=0.20\textwidth]{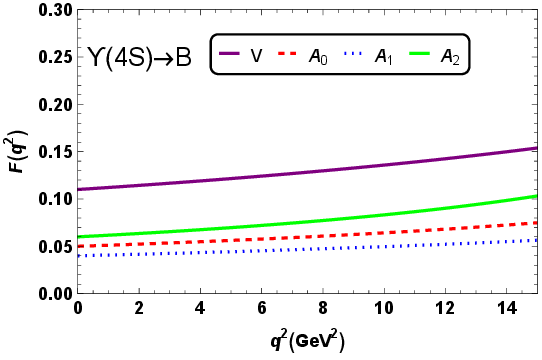}}\\
	\caption{The $q^2$ dependence of the $\Upsilon(nS)\to B_{(c)}$ transition form factors. }\label{T1}
\end{figure}

All computations are carried out within the $q^+=0$ reference frame, where the form factors can only be obtained at spacelike momentum transfers $q^2=-q^2_{\bot}\leq0$. However, the timelike region form factors are needed for physical decay processes. Here we use the
following double-pole approximation to parametrize the form factors in the spacelike region and then extend to the timelike region,
\be
F\left(q^{2}\right)=\frac{F(0)}{1-a q^{2} / m^{2}+b q^{4} / m^{4}},
\en
where $m$ represents the initial meson mass and $F(q^{2})$ denotes the different form factors $V,A_{0},A_{1}$ and $A_{2}$.
The values of $a$ and $b$ can be obtained by performing a 3-parameter fit to the form factors in the range $-15 \text{GeV}^2\leq q^2\leq0$, which are collected in Tables \ref{formfactor} and  \ref{formfactor2}. The uncertainties arise from the decay constants of the initial bottomonia ($\Upsilon(nS)$) and the final state mesons ($B, B_{c}$). Some of these form factors have been calculated by using other methods, such as the BSW model, the NRQCD and the BS equation method, which are listed in Table \ref{comp}.
One can find that our predictions are comparable to those theoretical results in most cases. 
In Fig. \ref{T1}, we also plot the $q^2$ dependence of the $\Upsilon(nS) \to B_{(c)}$ transition form factors. 

\begin{table}[H]
\caption{The $\Upsilon(nS)\to B_c$ transition form factors in the CLFQM. }
\begin{center}
\scalebox{1.0}{
\begin{tabular}{|c|c|c|c|c|}
\hline
  & $F_{i}(q^{2}=0)$&$F_{i}(q^{2}_{max})$&a&b\\
\hline
$V^{\Upsilon(1S)\rightarrow B_{c}}$&$1.18^{+0.00+0.01}_{-0.00-0.01}$&$1.47^{+0.00+0.01}_{-0.00-0.01}$&$1.86^{+0.01+0.02}_{-0.01-0.02}$ & $0.95^{+0.02+0.03}_{-0.02-0.02}$      \\
$A^{\Upsilon(1S)\rightarrow B_{c}}_{0}$&$0.39^{+0.01+0.01}_{-0.00-0.00}$&$0.46^{+0.01+0.01}_{-0.00-0.00}$&$1.40^{+0.00+0.02}_{-0.01-0.02}$& $0.51^{+0.01+0.02}_{-0.01-0.02}$      \\
$A^{\Upsilon(1S)\rightarrow B_{c}}_{1}$&$0.43^{+0.00+0.00}_{-0.00-0.00}$&$ 0.51^{+0.01+0.02}_{-0.02-0.02}$&$1.42^{+0.01+0.03}_{-0.01-0.01}$ & $0.51^{+0.00+0.00}_{-0.00-0.00}$        \\
$A^{\Upsilon(1S)\rightarrow B_{c}}_{2}$&$0.23^{+0.00+0.00}_{-0.01-0.00}$&$0.26^{+0.00+0.00}_{-0.01-0.00}$ &$1.13^{+0.02+0.04}_{-0.01-0.00}$ & $0.54^{+0.01+0.02}_{-0.01-0.00}$        \\
\hline
$V^{\Upsilon(2S)\rightarrow B_{c}}$&$0.57^{+0.02+0.01}_{-0.02-0.01}$&$0.68^{+0.04+0.01}_{-0.04-0.02}$&$1.20^{+0.14+0.01}_{-0.14-0.00}$ & $0.45^{+0.04+0.02}_{-0.02-0.03}$      \\
$A^{\Upsilon(2S)\rightarrow B_{c}}_{0}$&$0.28^{+0.01+0.00}_{-0.00-0.00}$&$0.34^{+0.01+0.00}_{-0.00-0.00}$&$1.35^{+0.04+0.01}_{-0.05-0.02}$ & $0.33^{+0.05+0.02}_{-0.05-0.02}$      \\
$A^{\Upsilon(2S)\rightarrow B_{c}}_{1}$&$0.26^{+0.00+0.00}_{-0.00-0.00}$&$0.31^{+0.00+0.03}_{-0.00-0.03}$&$1.12^{+0.07+0.01}_{-0.08-0.00}$ & $0.28^{+0.04+0.02}_{-0.03-0.02}$      \\
$A^{\Upsilon(2S)\rightarrow B_{c}}_{2}$&$0.38^{+0.02+0.00}_{-0.03-0.00}$&$0.50^{+0.04+0.02}_{-0.03-0.00}$&$2.04^{+0.00+0.04}_{-0.00-0.00}$ & $1.07^{+0.08+0.04}_{-0.07-0.00}$      \\
\hline
$V^{\Upsilon(3S)\rightarrow B_{c}}$&$2.22^{+0.10+0.01}_{-0.10-0.01}$&$3.36^{+0.15+0.00}_{-0.16-0.01}$&$2.91^{+0.02+0.04}_{-0.02-0.04}$ & $4.71^{+0.18+0.19}_{-0.19-0.17}$      \\
$A^{\Upsilon(3S)\rightarrow B_{c}}_{0}$&$0.37^{+0.03+0.00}_{-0.03-0.00}$&$0.49^{+0.04+0.00}_{-0.04-0.00}$&$2.22^{+0.00+0.04}_{-0.01-0.05}$& $4.06^{+0.17+0.23}_{-0.17-0.19}$      \\
$A^{\Upsilon(3S)\rightarrow B_{c}}_{1}$&$0.57^{+0.03+0.00}_{-0.04-0.01}$&$0.79^{+0.05+0.00}_{-0.05-0.01} $&$2.40^{+0.02+0.05}_{-0.02-0.04}$ & $3.77^{+0.17+0.19}_{-0.17-0.16}$        \\
$A^{\Upsilon(3S)\rightarrow B_{c}}_{2}$&$-0.42^{+0.02+0.01}_{-0.02-0.00}$&$-0.78^{+0.02+0.01}_{-0.02-0.00}$ &$3.32^{+0.02+0.03}_{-0.02-0.00}$ & $2.34^{+0.32+0.00}_{-0.35-0.02}$        \\
\hline
$V^{\Upsilon(4S)\rightarrow B_{c}}$&$0.36^{+0.02+0.01}_{-0.01-0.01}$&$0.42^{+0.03+0.01}_{-0.03-0.00}$&$0.92^{+0.15+0.01}_{-0.14-0.01}$ & $0.56^{+0.01+0.03}_{-0.00-0.03}$      \\
$A^{\Upsilon(4S)\rightarrow B_{c}}_{0}$&$0.20^{+0.01+0.00}_{-0.00-0.00}$&$0.26^{+0.01+0.00}_{-0.00-0.00}$&$1.37^{+0.07+0.01}_{-0.06-0.00}$& $0.22^{+0.05+0.02}_{-0.04-0.01}$      \\
$A^{\Upsilon(4S)\rightarrow B_{c}}_{1}$&$0.18^{+0.00+0.00}_{-0.00-0.00}$&$0.22^{+0.00+0.00}_{-0.01-0.00} $&$1.05^{+0.08+0.01}_{-0.08-0.00}$ & $0.24^{+0.03+0.01}_{-0.03-0.02}$        \\
$A^{\Upsilon(4S)\rightarrow B_{c}}_{2}$&$0.30^{+0.01+0.00}_{-0.01-0.00}$&$0.45^{+0.01+0.01}_{-0.01-0.00}$ &$2.16^{+0.05+0.02}_{-0.06-0.00}$ & $0.74^{+0.12+0.03}_{-0.10-0.00}$        \\
\hline
\end{tabular}\label{formfactor}
}
\end{center}
\end{table}

\begin{table}[H]
\caption{The $\Upsilon(nS)\to B$ transition form factors in the CLFQM,}
\begin{center}
\scalebox{1.0}{
\begin{tabular}{|c|c|c|c|c|}
\hline
  & $F_{i}(q^{2}=0)$&$F_{i}(q^{2}_{max})$&a&b\\
\hline
$V^{\Upsilon(1S)\rightarrow B}$&$0.20^{+0.00+0.07}_{-0.00-0.05}$&$0.39^{+0.00+0.13}_{-0.00-0.09}$&$3.61^{+0.03+0.31}_{-0.03-0.27}$ & $5.61^{+0.15+1.34}_{-0.15-1.03}$      \\
$A^{\Upsilon(1S)\rightarrow B}_{0}$&$0.06^{+0.00+0.02}_{-0.00-0.01}$&$0.10^{+0.00+0.03}_{-0.00-0.01}$&$2.89^{+0.02+0.34}_{-0.03-0.30}$ & $3.64^{+0.11+1.11}_{-0.10-0.81}$    \\
$A^{\Upsilon(1S)\rightarrow B}_{1}$&$0.06^{+0.00+0.02}_{-0.00-0.01}$&$0.11^{+0.00+0.03}_{-0.00-0.02}$&$2.93^{+0.02+0.32}_{-0.03-0.19}$ & $3.62^{+0.11+1.07}_{-0.11-0.79}$        \\
$A^{\Upsilon(1S)\rightarrow B}_{2}$&$0.06^{+0.00+0.00}_{-0.00-0.01}$&$0.10^{+0.00+0.00}_{-0.00-0.01}$&$2.74^{+0.02+0.042}_{-0.01-0.00}$ & $3.73^{+0.09+1.21}_{-0.09-0.00}$        \\
\hline
$V^{\Upsilon(2S)\rightarrow B}$&$0.15^{+0.01+0.03}_{-0.01-0.03}$&$0.20^{+0.00+0.03}_{-0.00-0.04}$&$2.39^{+0.16+0.56}_{-0.15-0.44}$ & $5.99^{+0.57+2.29}_{-0.49-1.53}$      \\
$A^{\Upsilon(2S)\rightarrow B}_{0}$&$0.06^{+0.00+0.01}_{-0.00-0.01}$&$0.10^{+0.00+0.02}_{-0.00-0.02}$&$2.91^{+0.12+0.39}_{-0.12-0.30}$& $4.92^{+0.51+2.00}_{-0.44-1.36}$      \\
$A^{\Upsilon(2S)\rightarrow B}_{1}$&$0.06^{+0.00+0.01}_{-0.01-0.02}$&$0.09^{+0.00+0.01}_{-0.02-0.03}$&$2.49^{+0.12+0.45}_{-0.12-0.35}$ & $4.58^{+0.47+1.91}_{-0.40-1.28}$        \\
$A^{\Upsilon(2S)\rightarrow B}_{2}$&$0.07^{+0.01+0.00}_{-0.00-0.01}$&$0.16^{+0.02+0.00}_{-0.00-0.04}$ &$4.02^{+0.16+0.23}_{-0.15-0.00}$ & $6.70^{+0.71+2.05}_{-0.61-0.00}$        \\
\hline
$V^{\Upsilon(3S)\rightarrow B}$&$0.20^{+0.01+0.12}_{-0.02-0.10}$&$0.03^{+0.00+0.03}_{-0.00-0.03}$&$11.84^{+0.63+4.69}_{-0.62-3.93}$ & $153.41^{+5.89+43.57}_{-5.94-38.20}$      \\
$A^{\Upsilon(3S)\rightarrow B}_{0}$&$0.01^{+0.00+0.01}_{-0.01-0.01}$&$0.00^{+0.01+0.01}_{-0.00-0.00}$&$0.93^{+0.16+0.40}_{-0.17-0.42}$& $14.25^{+5.89+43.57}_{-5.94-38.20}$      \\
$A^{\Upsilon(3S)\rightarrow B}_{1}$&$0.03^{+0.00+0.02}_{-0.01-0.02}$&$-0.01^{+0.00+0.00}_{-0.00-0.00} $&$2.00^{+0.18+0.81}_{-0.15-0.71}$ & $27.27^{+1.83+7.35}_{-1.49-6.79}$        \\
$A^{\Upsilon(3S)\rightarrow B}_{2}$&$-0.06^{+0.00+0.02}_{-0.00-0.00}$&$-0.14^{+0.14+0.11}_{-0.10-0.13}$ &$-2.38^{+0.94+1.07}_{-1.24-1.10}$ & $-25.88^{+20.70+22.32}_{-24.24-18.69}$        \\
\hline
$V^{\Upsilon(4S)\rightarrow B}$&$0.11^{+0.00+0.02}_{-0.00-0.02}$&$0.14^{+0.01+0.03}_{-0.00-0.03}$&$2.17^{+0.15+0.40}_{-0.15-0.32}$ & $4.83^{+0.57+2.08}_{-0.47-1.30}$      \\
$A^{\Upsilon(4S)\rightarrow B}_{0}$&$0.05^{+0.00+0.01}_{-0.00-0.01}$&$0.08^{+0.00+0.01}_{-0.00-0.02}$&$2.52^{+0.15+0.36}_{-0.13-0.26}$& $4.48^{+0.53+2.04}_{-0.45-1.30}$      \\
$A^{\Upsilon(4S)\rightarrow B}_{1}$&$0.04^{+0.01+0.01}_{-0.00-0.00}$&$ 0.06^{+0.01+0.00}_{-0.00-0.01}$&$2.21^{+0.14+0.37}_{-0.13-0.28}$ & $3.97^{+0.48+1.88}_{-0.40-1.18}$        \\
$A^{\Upsilon(4S)\rightarrow B}_{2}$&$0.06^{+0.01+0.00}_{-0.00-0.01}$&$0.10^{+0.02+0.00}_{-0.00-0.03}$ &$3.18^{+0.19+0.32}_{-0.17-0.00}$ & $6.33^{+0.76+2.41}_{-0.65-0.00}$        \\
\hline
\end{tabular}\label{formfactor2}
}
\end{center}
\end{table}

\begin{table}[H]
\caption{Comparison of the $\Upsilon(nS)\to B_{(c)}$ transition form factors at $q^2=0$ between this work and other literature.}
\begin{center}
\scalebox{1.2}{
\begin{tabular}{|c|c|c|c|c|c|}
\hline
&reference&$V$&$A_{0}$&$A_{1}$&$A_{2}$\\
\hline
$\Upsilon(1S)\to B_{c}$&This work&$1.18$&$0.39$&$0.43$&$0.23$\\
&BSW\cite{Dhir:2009rb}&$1.61$&$0.46$&$0.62$&$0.38$\\
&NRQCD\cite{Sun:2015nya}&$1.66$&$0.67$&$0.70$&$0.51$\\
&BS\cite{T. Wang}&$1.23$&$-$&$-$&$0.27$\\
\hline
$\Upsilon(2S) \to B_{c}$&This work&$0.57$&$0.28$&$0.26$&$0.38$\\
&NRQCD\cite{Sun:2015nya}&$1.44$&$0.65$&$0.69$&$0.48$\\
\hline
$\Upsilon(3S)\to B_{c}$&This work&$2.22$&$0.37$&$0.57$&$-0.42$\\
&NRQCD\cite{Sun:2015nya}&$1.25$&$0.57$&$0.64$&$0.29$\\
\hline
$\Upsilon(4S)\to B_{c}$&This work&$0.36$&$0.20$&$0.18$&$0.30$\\
\hline
$\Upsilon(1S)\to B$&This work&$0.20$&$0.06$&$0.06$&$0.06$\\
&BS\cite{T. Wang}&$0.38$&$-$&$-$&$0.013$\\
\hline
\end{tabular}\label{comp}
}
\end{center}
\end{table}

\subsection{Semileptonic decays $\Upsilon(nS) \to B_{(c)} \ell {\nu}_{\ell}$}
Using the input parameters and theoretical formulas provided above, we calculate the branching ratios of the semileptonic decays $\Upsilon(nS) \to B_{(c)} \ell {\nu}_{\ell}$, which are summarized in Tables \ref{B_c1}-\ref{B1}, where the uncertainties arise from the initial bottomonium full widths, the decay constants of initial and final state mesons, respectively. Other relevant theoretical results based on the BSW model \cite{Dhir:2009rb}, NRQCD \cite{Chang:2016gyw},  CCQM \cite{Tran:2024phq} and BS equation \cite{T. Wang} are also listed for comparison. In Table \ref{B_c1}, our predictions for the branching ratios of the decays $\Upsilon(1S)\to B_{c}\ell\nu_{\ell}$ are consistent with the BSW, CCQM and BS results, while much smaller than the NRQCD calculations. It is strange that the branching ratios of the decays $\Upsilon(3S)\to B_{c}\ell\nu_{\ell}$ agree with NRQCD calculations, while those of the decays $\Upsilon(2S)\to B_{c}\ell\nu_{\ell}$ are at least one order magnitude smaller than those of NRQCD. Furthermore, there also exists an obvious difference in the branching ratios of the decays $\Upsilon(1S)\to B\ell\nu_{\ell}$ between our results and BS equation calculations, which is shown in Table \ref{B1}. It needs more theoretical effort to clarify these divergences. The $\Upsilon(4S)$ has a much larger decay width than those of the other three states, which results in very tiny branching ratios for the decays $\Upsilon(4S) \to B_{(c)}\ell\nu_{\ell}$.

\begin{table}[H]
\caption{The branching ratios of the decays $\Upsilon(nS)\to B_{c}\ell\nu_{\ell}$ in the CLFQM with other theoretical results for comparison. }
\begin{center}
\scalebox{0.75}{
\begin{tabular}{|c |c |c |c|}
\hline
&$10^{-10}\times\mathcal{B} r(\Upsilon(1S)\to B_{c}e\nu_{e})$&$10^{-10}\times\mathcal{B} r(\Upsilon(1S)\to B_{c}\mu\nu_{\mu})$&$10^{-10}\times\mathcal{B} r(\Upsilon(1S)\to B_{c}\tau\nu_{\tau})$\\
\hline
This work&$2.10^{+0.00+0.01+0.05}_{-0.01-0.01-0.05}$&$2.08^{+0.00+0.01+0.05}_{-0.01-0.01-0.05}$&$0.35^{+0.00+0.00+0.01}_{-0.00-0.00-0.01}$\\
BSW\cite{Dhir:2009rb}&$1.70^{+0.03}_{-0.02}$&$1.69^{+0.04}_{-0.02}$&$0.29^{+0.05}_{-0.02}$\\
NRQCD\cite{Chang:2016gyw}&$5.58^{+3.32+0.14+0.08}_{-1.54-0.12-0.18}$&$5.58^{+3.32+0.14+0.08}_{-1.54-0.12-0.18}$&$1.30^{+0.77+0.03+0.02}_{-0.35-0.03-0.04}$\\
CCQM\cite{Tran:2024phq}&$1.84$&$1.83$&$0.47$\\
BS\cite{T. Wang}&$1.37^{+0.22}_{-0.19}$&$1.37^{+0.22}_{-0.19}$&$0.42^{+0.06}_{-0.05}$\\
\hline
&$10^{-10}\times\mathcal{B} r(\Upsilon(2S)\to B_{c}e\nu_{e})$&$10^{-10}\times\mathcal{B} r(\Upsilon(2S)\to B_{c}\mu\nu_{\mu})$&$10^{-10}\times\mathcal{B} r(\Upsilon(2S)\to B_{c}\tau\nu_{\tau})$\\
\hline
This work&$1.96^{+0.07+0.00+0.18}_{-0.10-0.00-0.15}$&$1.96^{+0.10+0.00+0.18}_{-0.07-0.00-0.15}$&$0.64^{+0.02+0.00+0.06}_{-0.01-0.00-0.05}$\\
NRQCD\cite{Chang:2016gyw}&$21.4^{+12.6+2.0+0.4}_{-5.8-1.6-0.6}$&$21.4^{+12.6+2.0+0.4}_{-5.8-1.6-0.6}$&$8.08^{+4.78+0.72+0.12}_{-2.22-0.62-0.26}$\\
\hline
&$10^{-9}\times\mathcal{B} r(\Upsilon(3S)\to B_{c}e\nu_{e})$&$10^{-9}\times\mathcal{B} r(\Upsilon(3S)\to B_{c}\mu\nu_{\mu})$&$10^{-9}\times\mathcal{B} r(\Upsilon(3S)\to B_{c}\tau\nu_{\tau})$\\
\hline
This work&$5.01^{+0.43+0.15+0.50}_{-0.56-0.02-0.42}$&$4.97^{+0.43+0.15+0.50}_{-0.56-0.02-0.41}$&$1.48^{+0.14+0.04+0.15}_{-0.18-0.01-0.12}$\\
NRQCD\cite{Chang:2016gyw}&$4.48^{+2.66+0.44+0.08}_{-1.22-0.38-0.14}$&$4.48^{+2.66+0.44+0.08}_{-1.22-0.38-0.14}$&$2.06^{+1.22+0.20+0.04}_{-0.56-0.18-0.06}$\\
\hline
&$10^{-13}\times\mathcal{B} r(\Upsilon(4S)\to B_{c}e\nu_{e})$&$10^{-13}\times\mathcal{B} r(\Upsilon(4S)\to B_{c}\mu\nu_{\mu})$&$10^{-13}\times\mathcal{B} r(\Upsilon(4S)\to B_{c}\tau\nu_{\tau})$\\
\hline
This work&$2.39^{+0.09+0.00+0.31}_{-0.09-0.00-0.26}$&$2.39^{+0.09+0.00+0.33}_{-0.08-0.00-0.26}$&$1.13^{+0.04+0.00+0.16}_{-0.01-0.00-0.12}$\\
\hline
\end{tabular}\label{B_c1}}
\end{center}
\end{table}

\begin{table}[H]
	\caption{The branching ratios of the decays $\Upsilon(nS)\to B\ell\nu_{\ell}$ in the CLFQM with other theoretical results for comparison. }
	\begin{center}
		\scalebox{0.9}{
			\begin{tabular}{|c |c |c |c|}
				\hline
				&$10^{-13}\times\mathcal{B} r(\Upsilon(1S)\to Be\nu_{e})$&$10^{-13}\times\mathcal{B} r(\Upsilon(1S)\to B\mu\nu_{\mu})$&$10^{-13}\times\mathcal{B} r(\Upsilon(1S)\to B\tau\nu_{\tau})$\\
				\hline
				This work&$1.39^{+0.00+1.16+0.03}_{-0.56-0.39-0.03}$&$1.38^{+0.00+1.15+0.03}_{-0.55-0.38-0.03}$&$0.60^{+0.01+0.44+0.01}_{-0.10-0.16-0.01}$\\
				BS\cite{T. Wang}&$7.83^{+1.40}_{-1.20}$&$7.82^{+1.40}_{-1.20}$&$5.04^{+0.92}_{-0.79}$\\
				\hline
				&$10^{-13}\times\mathcal{B} r(\Upsilon(2S)\to Be\nu_{e})$&$10^{-13}\times\mathcal{B} r(\Upsilon(2S)\to B\mu\nu_{\mu})$&$10^{-13}\times\mathcal{B} r(\Upsilon(2S)\to B\tau\nu_{\tau})$\\
				\hline
				This work&$3.09^{+0.29+1.25+0.28}_{-1.14-1.83-0.23}$&$3.08^{+0.29+1.24+0.28}_{-1.13-1.83-0.23}$&$1.63^{+0.12+0.57+0.15}_{-0.49-0.87-0.12}$\\
				\hline
				&$10^{-13}\times\mathcal{B} r(\Upsilon(3S)\to Be\nu_{e})$&$10^{-13}\times\mathcal{B} r(\Upsilon(3S)\to B\mu\nu_{\mu})$&$10^{-13}\times\mathcal{B} r(\Upsilon(3S)\to B\tau\nu_{\tau})$\\
				\hline
				This work&$4.07^{+0.15+5.16+0.41}_{-1.54-3.23-0.22}$&$4.03^{+0.15+5.11+0.40}_{-1.49-3.20-0.22}$&$1.18^{+0.07+1.60+0.12}_{-0.44-0.94-0.10}$\\
				\hline
				&$10^{-16}\times\mathcal{B} r(\Upsilon(4S)\to Be\nu_{e})$&$10^{-16}\times\mathcal{B} r(\Upsilon(4S)\to B\mu\nu_{\mu})$&$10^{-16}\times\mathcal{B} r(\Upsilon(4S)\to B\tau\nu_{\tau})$\\
				\hline
				This work&$2.91^{+0.05+0.50+0.40}_{-1.75-2.24-0.32}$&$2.91^{+0.05+0.49+0.40}_{-1.74-2.24-0.32}$&$2.01^{+0.04+0.07+0.28}_{-0.91-1.28-0.22}$\\
				\hline
			\end{tabular}\label{B1}}
	\end{center}
\end{table}
In Table \ref{R},  we also calculate the lepton flavor universality (LFU) ratios, which are defined as
\be
R^{B_{(c)}}_{\Upsilon(nS)}=\frac{\Gamma\left(\Upsilon(nS) \rightarrow B_{(c)} \tau \nu_\tau\right)}{\Gamma\left(\Upsilon(nS)\rightarrow B_{(c)} \ell^{\prime} \nu_{\ell^{\prime}}\right)},
\en
where $\ell^{\prime}$ refers to $e, \mu$ and a large part of the theoretical and experimental uncertainties, especially the errors from the form factors, can be canceled.  One can find that most of our
predictions are comparable with other theoretical results from the BSW model \cite{Dhir:2009rb}, BS equation \cite{T. Wang}, CCQM \cite{Tran:2024phq} and NRQCD \cite{Chang:2016gyw}.
\begin{table}[H]
	\caption{The LUF ratios $R^{B_{(c)}}_{\Upsilon(nS)}$ in the CLFQM with the reults from the BSW model \cite{Dhir:2009rb}, BS equation \cite{T. Wang}, CCQM \cite{Tran:2024phq} and NRQCD \cite{Chang:2016gyw} for comparsion.}
	\begin{center}
		\scalebox{0.9}{
			\begin{tabular}{c|c|c|c}
				\hline\hline
				Ratios&Predicted values&Ratios&Predicted values\\
				\hline
				$\multirow{2}{*}{$\displaystyle R^{B_{c}}_{\Upsilon(1S)}=\frac{\Gamma\left(\Upsilon(1S)\to B_{c}\tau\nu_{\tau}\right)}{\Gamma\left(\Upsilon(1S)\to B_{c}\ell^{\prime}\nu_{\ell^{\prime}}\right)}$}$&$0.167^{+0.001+0.001+0.000}_{-0.001-0.001-0.001}$&$\multirow{2}{*}{$\displaystyle R^{B}_{\Upsilon(1S)}=\frac{\Gamma\left(\Upsilon(1S)\to B\tau\nu_{\tau}\right)}{\Gamma\left(\Upsilon(1S)\to B\ell^{\prime}\nu_{\ell^{\prime}}\right)}$}$&$0.432^{+0.050+0.008+0.002}_{-0.000-0.024-0.002}$\\
                &$0.24$ \cite{Chang:2016gyw};\;\;$0.30$\cite{T. Wang}&&$0.64\cite{T. Wang}$\\
             &$0.17$\cite{Dhir:2009rb};\;\;$0.26$\cite{Tran:2024phq}&&$0.55$\cite{Tran:2024phq}\\
                \hline
				$\multirow{2}{*}{$\displaystyle R^{B_{c}}_{\Upsilon(2S)}=\frac{\Gamma\left(\Upsilon(2S)\to B_{c}\tau\nu_{\tau}\right)}{\Gamma\left(\Upsilon(2S)\to B_{c}\ell^{\prime}\nu_{\ell^{\prime}}\right)}$}$&$0.327^{+0.012+0.000+0.025}_{-0.002-0.001-0.030}$&$\multirow{2}{*}{$\displaystyle R^{B}_{\Upsilon(2S)}=\frac{\Gamma\left(\Upsilon(2S)\to B\tau\nu_{\tau}\right)}{\Gamma\left(\Upsilon(2S)\to B\ell^{\prime}\nu_{\ell^{\prime}}\right)}$}$&$0.528^{+0.057+0.075+0.000}_{-0.010-0.021-0.000}$\\
                &$0.38^{+0.03}_{-0.01}\cite{Chang:2016gyw}$&&$$\\
                \hline
               $\multirow{2}{*}{$\displaystyle R^{B_{c}}_{\Upsilon(3S)}=\frac{\Gamma\left(\Upsilon(3S)\to B_{c}\tau\nu_{\tau}\right)}{\Gamma\left(\Upsilon(3S)\to B_{c}\ell^{\prime}\nu_{\ell^{\prime}}\right)}$}$&$0.295^{+0.003+0.000+0.001}_{-0.003-0.000-0.000}$&$\multirow{2}{*}{$\displaystyle R^{B}_{\Upsilon(3S)}=\frac{\Gamma\left(\Upsilon(3S)\to B\tau\nu_{\tau}\right)}{\Gamma\left(\Upsilon(3S)\to B\ell^{\prime}\nu_{\ell^{\prime}}\right)}$}$&$0.290^{+0.014+0.021+0.000}_{-0.003-0.003-0.001}$\\
                &$0.46^{+0.03}_{-0.01}\cite{Chang:2016gyw}$&&$$\\
                \hline
                $R^{B_c}_{\Upsilon(4S)}=\frac{\Gamma\left(\Upsilon(4S)\to B_{c}\tau\nu_{\tau}\right)}{\Gamma\left(\Upsilon(4S)\to B_{c}\ell^{\prime}\nu_{\ell^{\prime}}\right)}$&$0.473^{+0.014+0.000+0.005}_{-0.001-0.000-0.000}$&$R^{B}_{\Upsilon(4S)}=\frac{\Gamma\left(\Upsilon(4S)\to B\tau\nu_{\tau}\right)}{\Gamma\left(\Upsilon(4S)\to B\ell^{\prime}\nu_{\ell^{\prime}}\right)}$&$0.693^{+0.001+0.115+0.000}_{-0.001-0.053-0.001}$\\
				\hline\hline
			\end{tabular}
			\label{R}
			}
	\end{center}
\end{table}  

\subsection{Physical observables $A_{FB}$ and $f_L$}
\begin{figure}[H]
	\vspace{0.32cm}
	\centering
	\subfigure[]{\includegraphics[width=0.22\textwidth]{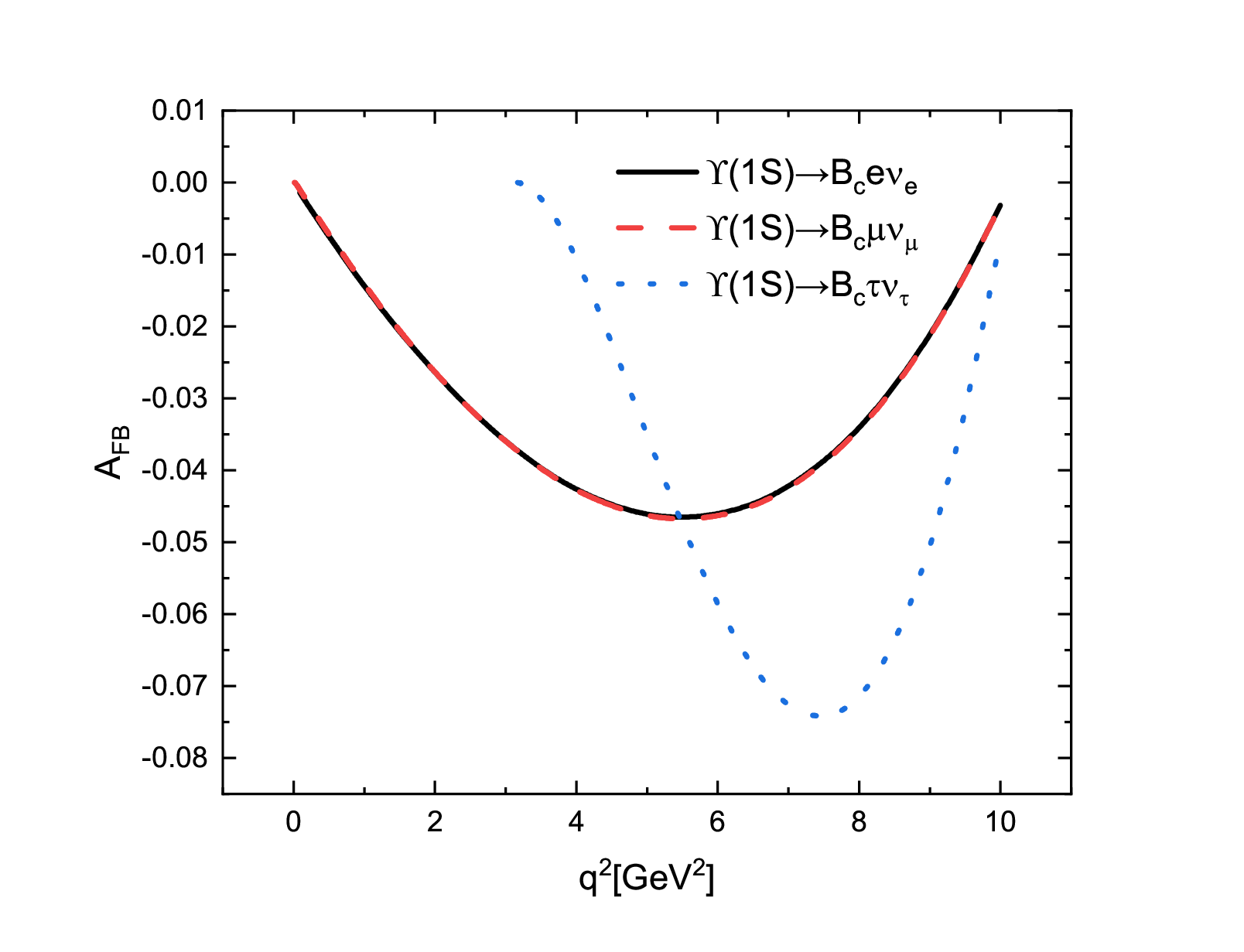}}
	\subfigure[]{\includegraphics[width=0.22\textwidth]{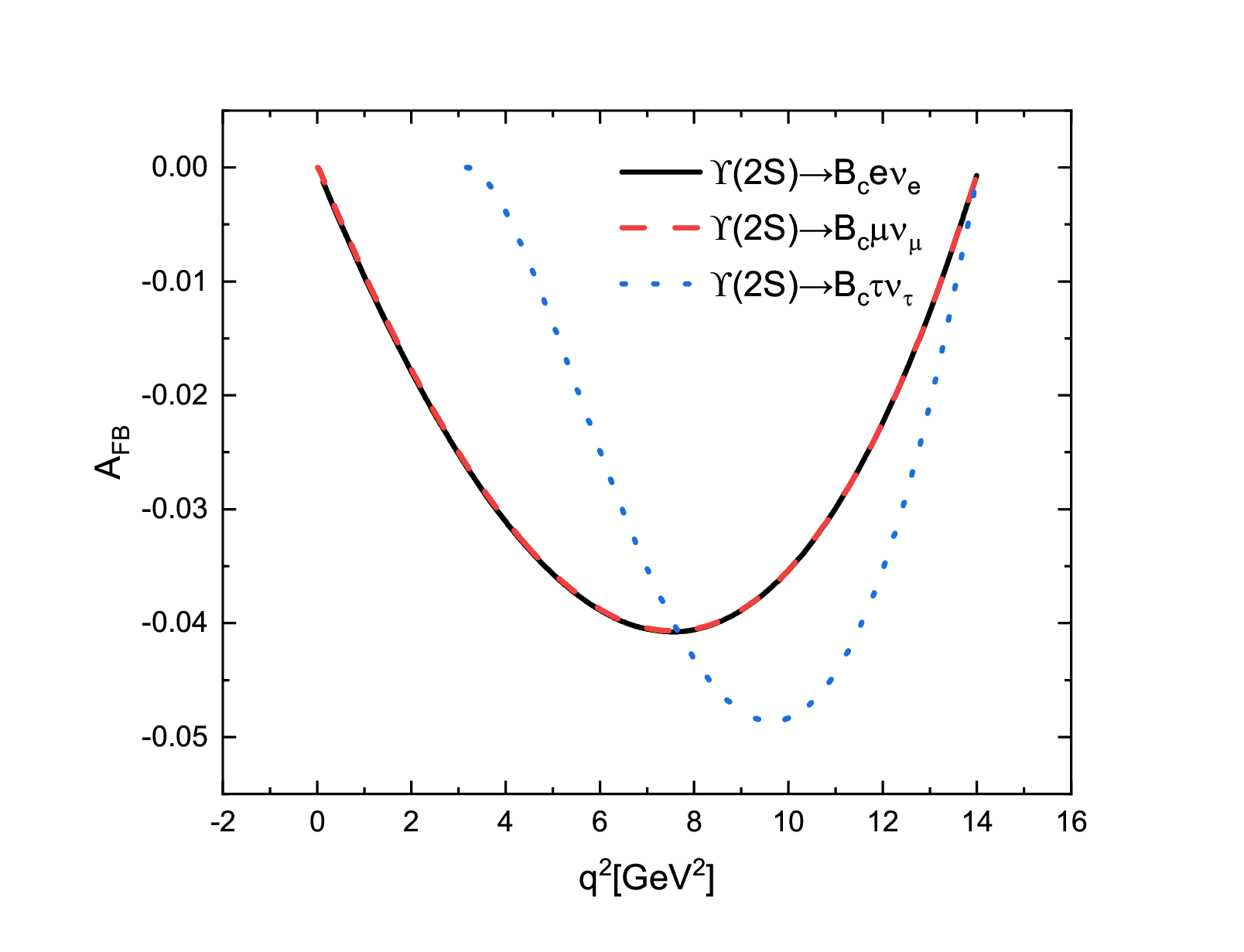}}
	\subfigure[]{\includegraphics[width=0.22\textwidth]{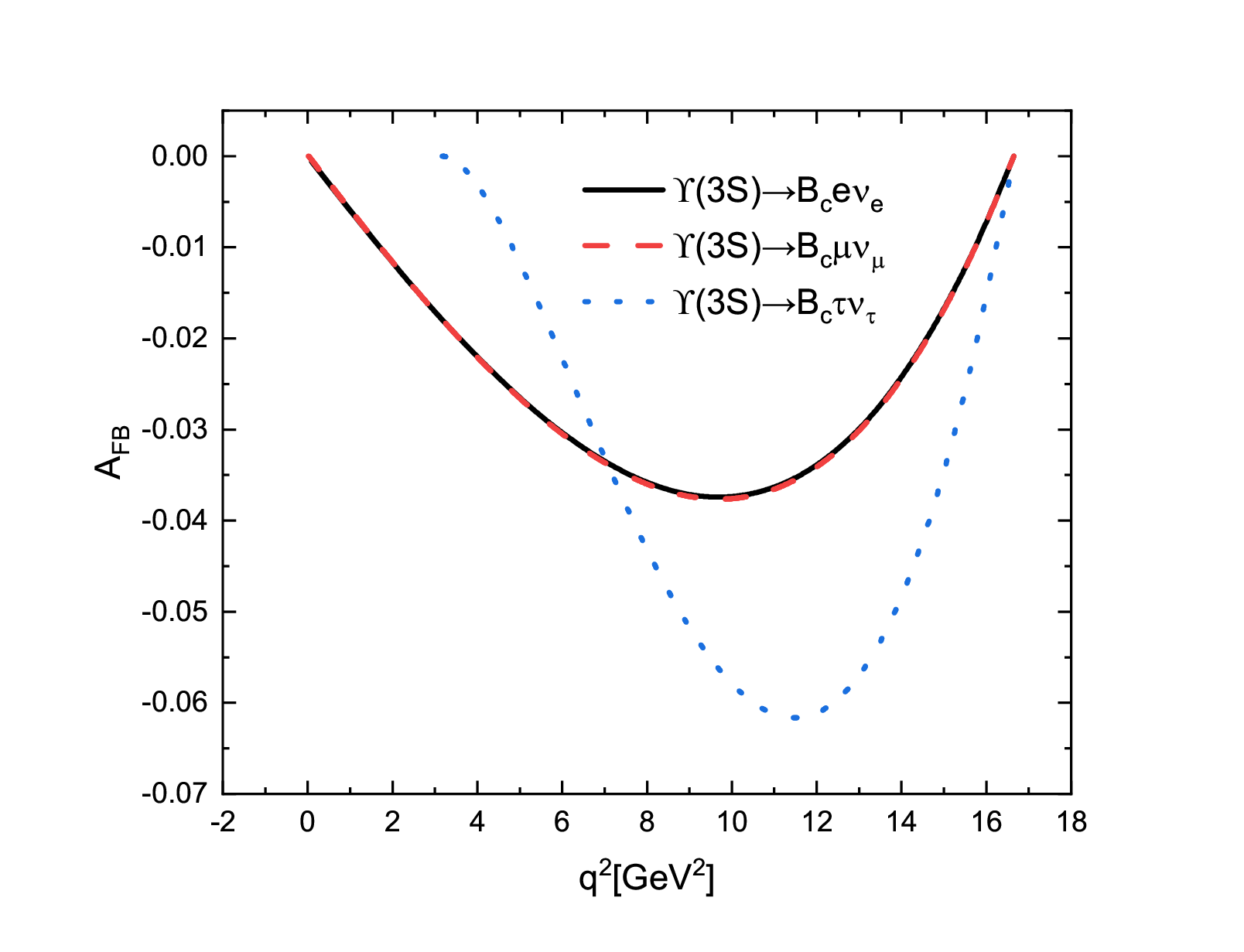}}
	\subfigure[]{\includegraphics[width=0.22\textwidth]{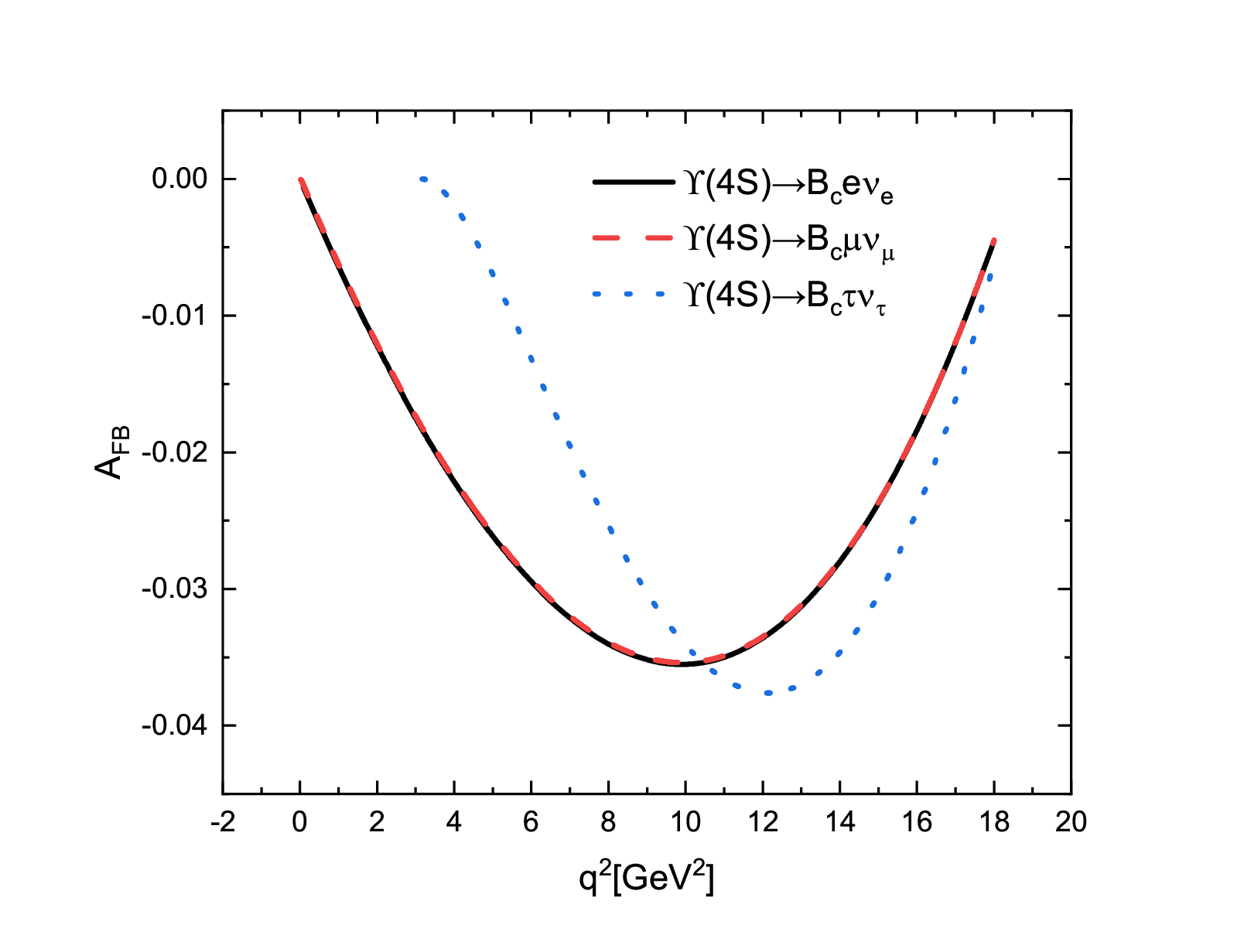}}\\
    \subfigure[]{\includegraphics[width=0.22\textwidth]{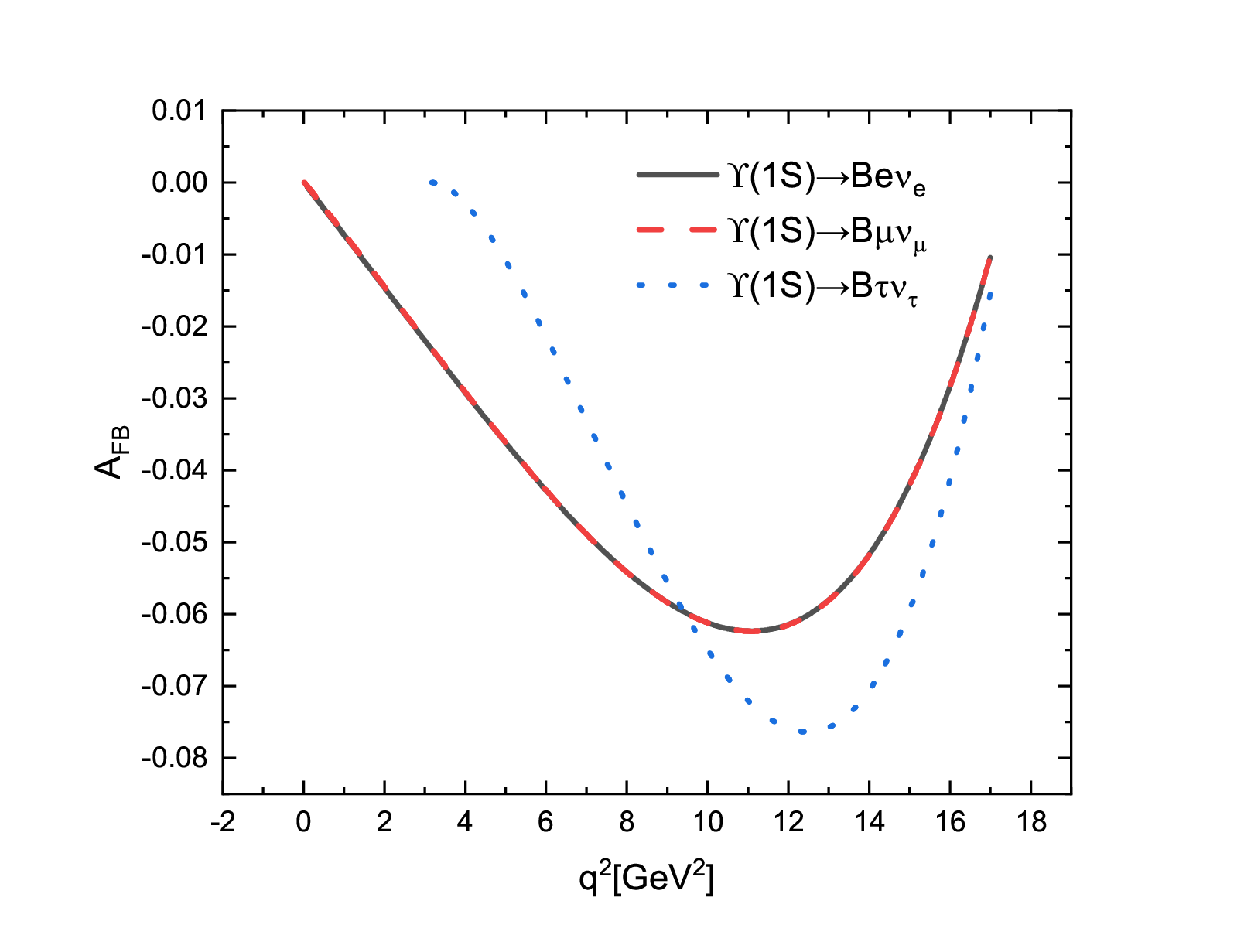}}
	\subfigure[]{\includegraphics[width=0.22\textwidth]{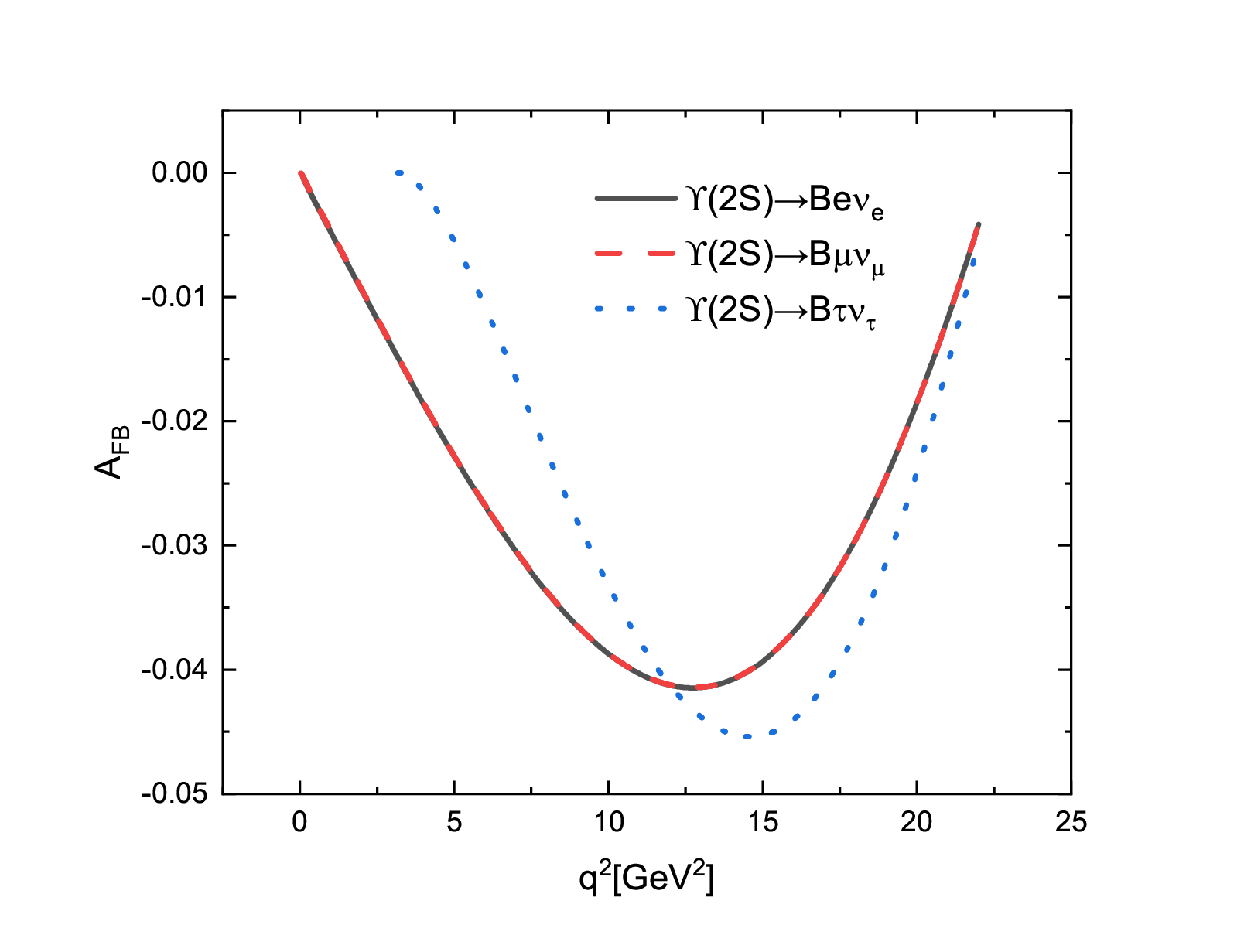}}
	\subfigure[]{\includegraphics[width=0.22\textwidth]{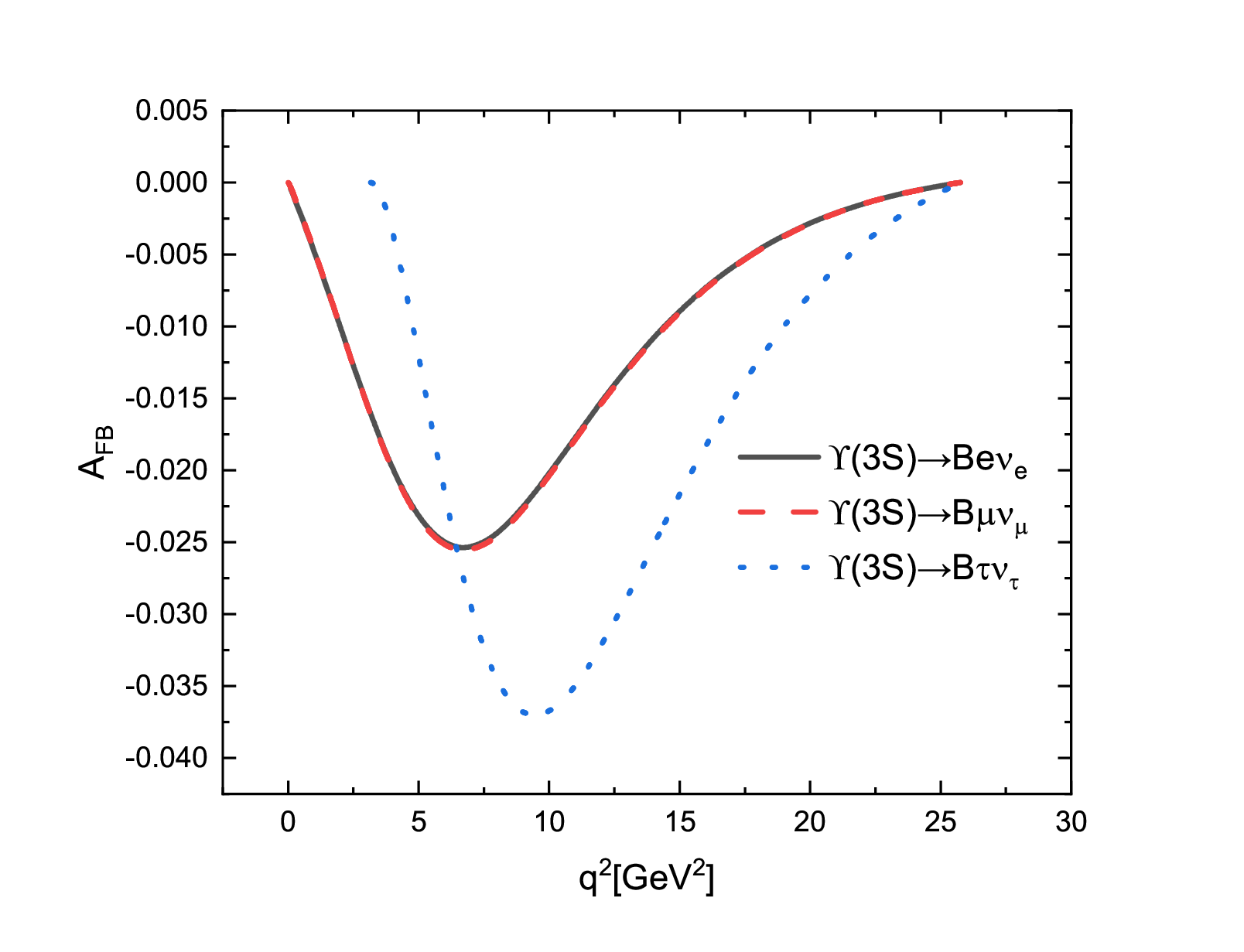}}
	\subfigure[]{\includegraphics[width=0.22\textwidth]{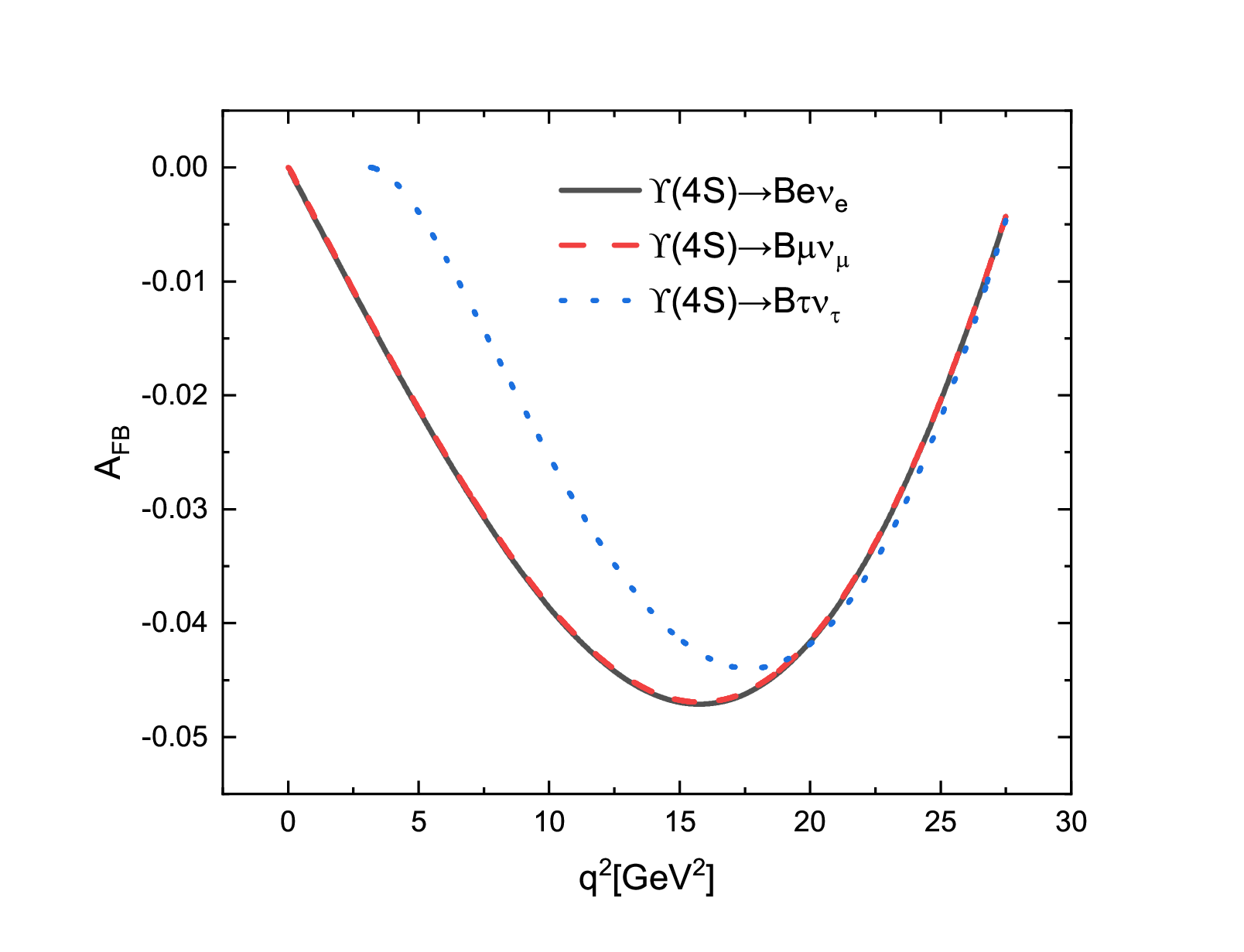}}
	\caption{The $q^2$ dependence of the forward-backward asymmetries
		$A_{FB}$ for the decays $\Upsilon(nS) \to B_{(c)} \ell {\nu}_{\ell}$. }\label{fig:T1}
\end{figure}

From Table \ref{AFB}, one can find that the $A_{FB}$ values for these semileptonic decays $\Upsilon(1S, 2S, 4S)\to B_{(c)}\ell\nu_\ell$ exhibit an obvious regularity, that is $A_{FB}(\Upsilon(nS)\to B \ell\nu_\ell)$ are about two times of the corresponding $A_{FB}(\Upsilon(nS)\to B_c \ell\nu_\ell)$ in magnitude. While there is an anomaly in the semileptonic decays $\Upsilon(3S)\to B_{(c)}\ell\nu_\ell$, where the sizes of the $A_{FB}(\Upsilon(3S)\to B \ell\nu_\ell)$ are smaller than those of the corresponding $A_{FB}(\Upsilon(3S)\to B_c \ell\nu_\ell)$. Another abnormal phenomenon is that the sizes of the $A_{FB}(\Upsilon(3S)\to B_{(c)} \tau\nu_\tau)$ are larger than those of the corresponding $A_{FB}(\Upsilon(3S)\to B_{(c)} \ell^\prime\nu_{\ell^\prime})$. It is contrary to the cases of other $\Upsilon(nS)$ decays. This should be connected with the $\Upsilon(3S)\to B_{(c)}$ transition form factors. We hope these abnormal results can be clarified by further experimental and theoretical studies.
\begin{table}[H]
\caption{The forward-backward asymmetries $A_{FB}$ for the decays $\Upsilon(nS)\to B_{(c)} \ell\nu_{\ell}$.}
\begin{center}
\scalebox{0.6}{
\begin{tabular}{|c|c|c|c||c|c|c|}
\hline\hline
 Channel  &$\Upsilon(1S)\to B_c e\nu_{e}$&$\Upsilon(1S)\to B_c \mu^{+}\nu_{\mu}$&$\Upsilon(1S)\to B_c \tau\nu_{\tau}$&$\Upsilon(1S)\to Be\nu_{e}$&$\Upsilon(1S)\to B \mu\nu_{\mu}$&$\Upsilon(1S)\to B \tau\nu_{\tau}$\\
 \hline
 $A_{FB}$&$-0.315^{+0.000+0.002+0.007}_{-0.000-0.002-0.007}$&$-0.315^{+0.000+0.002+0.007}_{-0.000-0.002-0.007}$&$-0.310^{+0.001+0.001+0.006}_{-0.001-0.002-0.007}$&$-0.690^{+0.002+0.235+0.016}_{-0.111-0.489-0.016}$&$-0.690^{+0.002+0.235+0.016}_{-0.111-0.488-0.016}$&$-0.644^{+0.002+0.218+0.015}_{-0.102-0.447-0.015}$\\
 \hline
  Channel  &$\Upsilon(2S)\to B_c e\nu_{e}$&$\Upsilon(2S)\to B_c \mu\nu_{\mu}$&$\Upsilon(2S)\to B_c \tau\nu_{\tau}$&$\Upsilon(2S)\to Be\nu_{e}$&$\Upsilon(2S)\to B \mu\nu_{\mu}$&$\Upsilon(2S)\to B \tau\nu_{\tau}$\\
   \hline
 $A_{FB}$&$-0.386^{+0.020+0.007+0.029}_{-0.020-0.007-0.035}$&$-0.385^{+0.020+0.007+0.029}_{-0.020-0.007-0.034}$&$-0.322^{+0.019+0.006+0.024}_{-0.019-0.007-0.029}$&$-0.599^{+0.089+0.258+0.045}_{-0.026-0.191-0.054}$&$-0.598^{+0.089+0.258+0.045}_{-0.026-0.191-0.054}$&$-0.530^{+0.078+0.227+0.040}_{-0.023-0.165-0.048}$\\
\hline\hline
 Channel  &$\Upsilon(3S)\to B_c e\nu_{e}$&$\Upsilon(3S)\to B_c \mu\nu_{\mu}$&$\Upsilon(3S)\to B_c \tau\nu_{\tau}$&$\Upsilon(3S)\to Be\nu_{e}$&$\Upsilon(3S)\to B \mu\nu_{\mu}$&$\Upsilon(3S)\to B \tau\nu_{\tau}$\\
\hline
 $A_{FB}$&$-0.403^{+0.045+0.000+0.034}_{-0.040-0.007-0.040}$&$-0.405^{+0.047+0.000+0.035}_{-0.042-0.007-0.040}$&$-0.497^{+0.055+0.001+0.041}_{-0.049-0.007-0.050}$&$-0.265^{+0.063+0.208+0.043}_{-0.089-0.674-0.027}$&$-0.267^{+0.035+0.209+0.044}_{-0.089-0.678-0.027}$&$-0.323^{+0.040+0.237+0.032}_{-0.160-0.887-0.103}$\\
\hline
 Channel  &$\Upsilon(4S)\to B_c e\nu_{e}$&$\Upsilon(4S)\to B_c \mu\nu_{\mu}$&$\Upsilon(4S)\to B_c \tau\nu_{\tau}$&$\Upsilon(4S)\to Be\nu_{e}$&$\Upsilon(4S)\to B \mu\nu_{\mu}$&$\Upsilon(4S)\to B \tau\nu_{\tau}$\\
\hline
 $A_{FB}$&$-0.447^{+0.021+0.000+0.049}_{-0.035-0.013-0.062}$&$-0.445^{+0.021+0.000+0.048}_{-0.035-0.013-0.062}$&$-0.358^{+0.020+0.000+0.039}_{-0.031-0.011-0.050}$&$-0.848^{+0.018+0.136+0.092}_{-0.187-0.361-0.118}$&$-0.844^{+0.018+0.136+0.092}_{-0.186-0.360-0.117}$&$-0.667^{+0.015+0.110+0.073}_{-0.145-0.281-0.093}$\\
\hline\hline
\end{tabular}\label{AFB}}
\end{center}
\end{table}
In Figure \ref{fig:T1}, we plot the $q^2$ dependence of the forward-backward asymmetries for the decays $\Upsilon(nS) \to B_{(c)} \ell \nu_{\ell}$.  The decays $\Upsilon(1S, 2S, 4S)\to B\ell\nu_\ell$ have larger $q^2$ ranges compared to those of the decays $\Upsilon(1S, 2S, 4S)\to B_{c}\ell\nu_\ell$, meanwhile similar $A_{FB}$ extrema between the corresponding decays, which results in the $A_{FB}$ values of the former are much larger than those of the latter. As for the decays $\Upsilon(3S)\to B\ell\nu_\ell$, the advantage of the larger $q^2$ range is canceled out by the smaller 
$A_{FB}$ extrema compared to those of the decays $\Upsilon(3S)\to B_c\ell\nu_\ell$, so the $A_{FB}$ values between the decays $\Upsilon(3S)\to B\ell\nu_\ell$ and  $\Upsilon(3S)\to B_c\ell\nu_\ell$ are close to each other.

In Table {\ref{FL}}, we can clearly find that the longitudinal polarization fractions $f_L$ between
the decays $\Upsilon(nS)\to B_{(c)}e\nu_{e}$ and $\Upsilon(nS)\to B_{(c)}\mu\nu_{\mu}$ are very close to each other, 
which reflects the LFU. In order to investigate the $q^2$ dependence of the polarization, we divide the full energy region into two segments for each decay and calculate the longitudinal polarization fractions accordingly. Region 1 is defined as $m_{\ell}^{2}<q^{2}<\frac{(m_{\Upsilon}-m_{B_{(c)}})^{2}+m_{\ell}^{2}}{2} $ and Region 2 is $\frac{(m_{\Upsilon}-m_{B_{(c)}})^{2}+m_{\ell}^{2}}{2} <q^{2}<(m_{\Upsilon}-m_{B_{(c)}})^{2}$.  Obviously, the longitudinal polarization in both regions for each decay is dominant; Furthermore, the longitudinal polarization fractions of the decays $\Upsilon(nS)\to B\ell\nu_{\ell}$ are a little larger than those of the corresponding decays $\Upsilon(nS)\to B_c\ell\nu_{\ell}$.  These results can be validated by future high-luminosity experiments.

\begin{figure}[H]
	\vspace{0.32cm}
	\centering
	\subfigure[]{\includegraphics[width=0.22\textwidth]{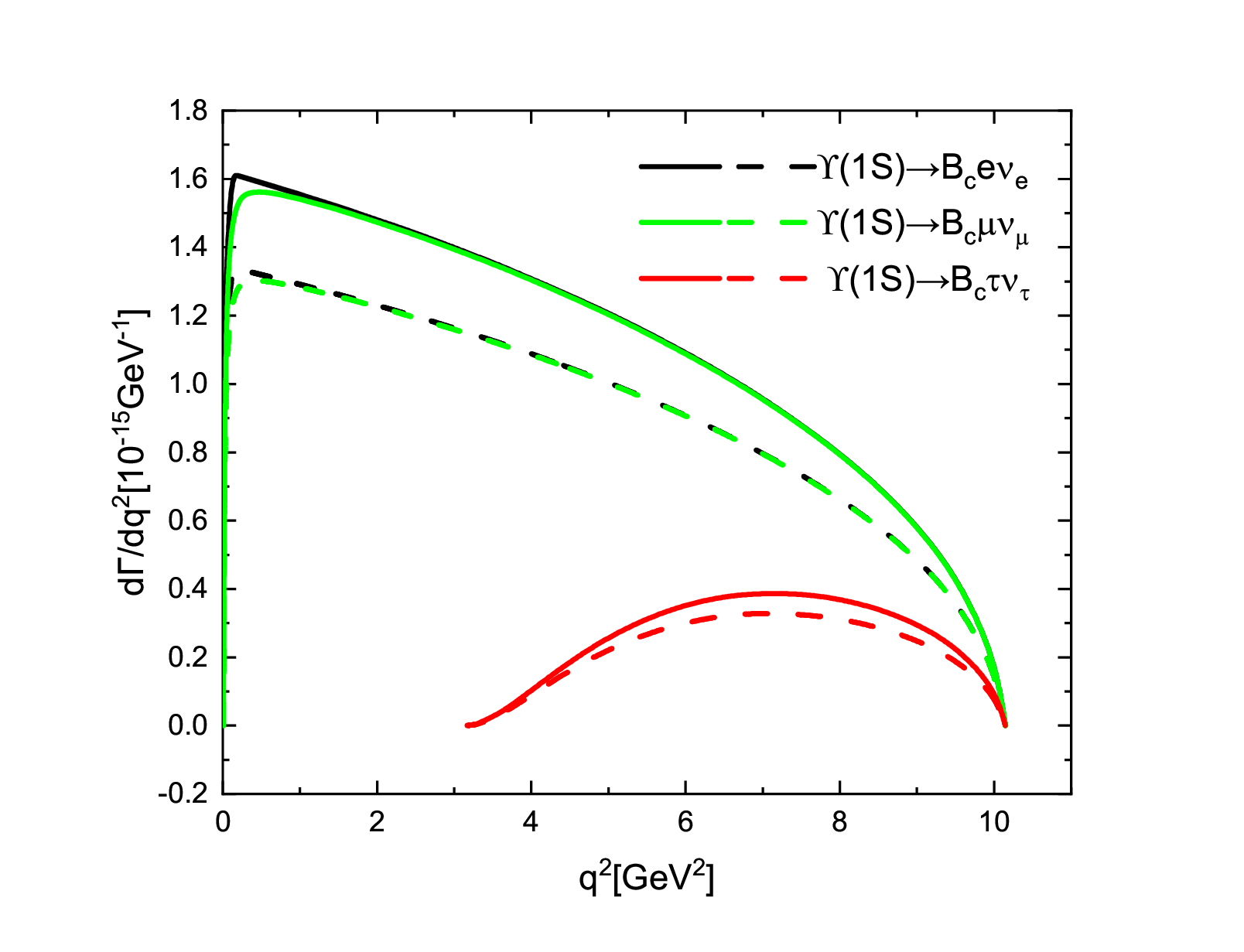}}
	\subfigure[]{\includegraphics[width=0.22\textwidth]{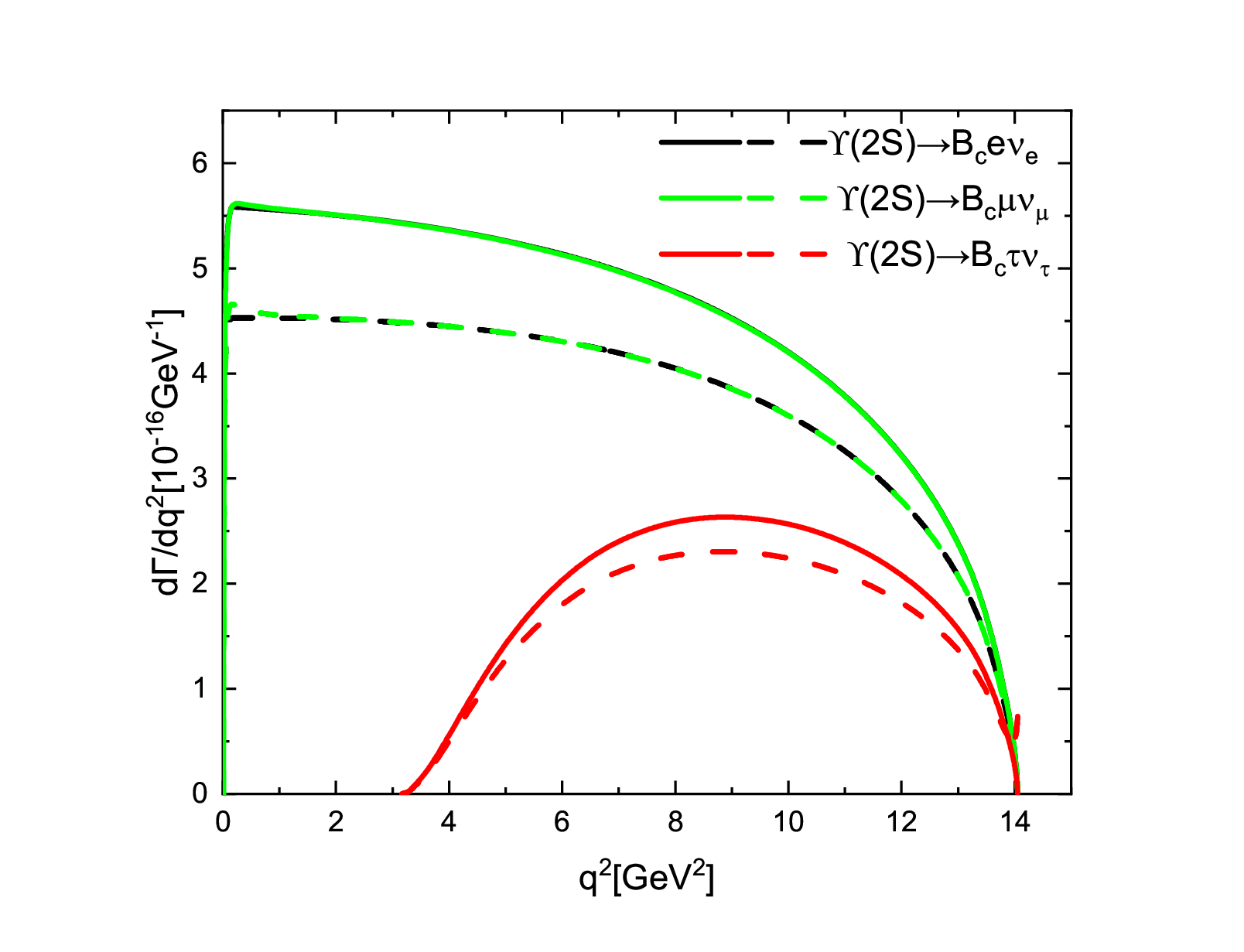}}
	\subfigure[]{\includegraphics[width=0.22\textwidth]{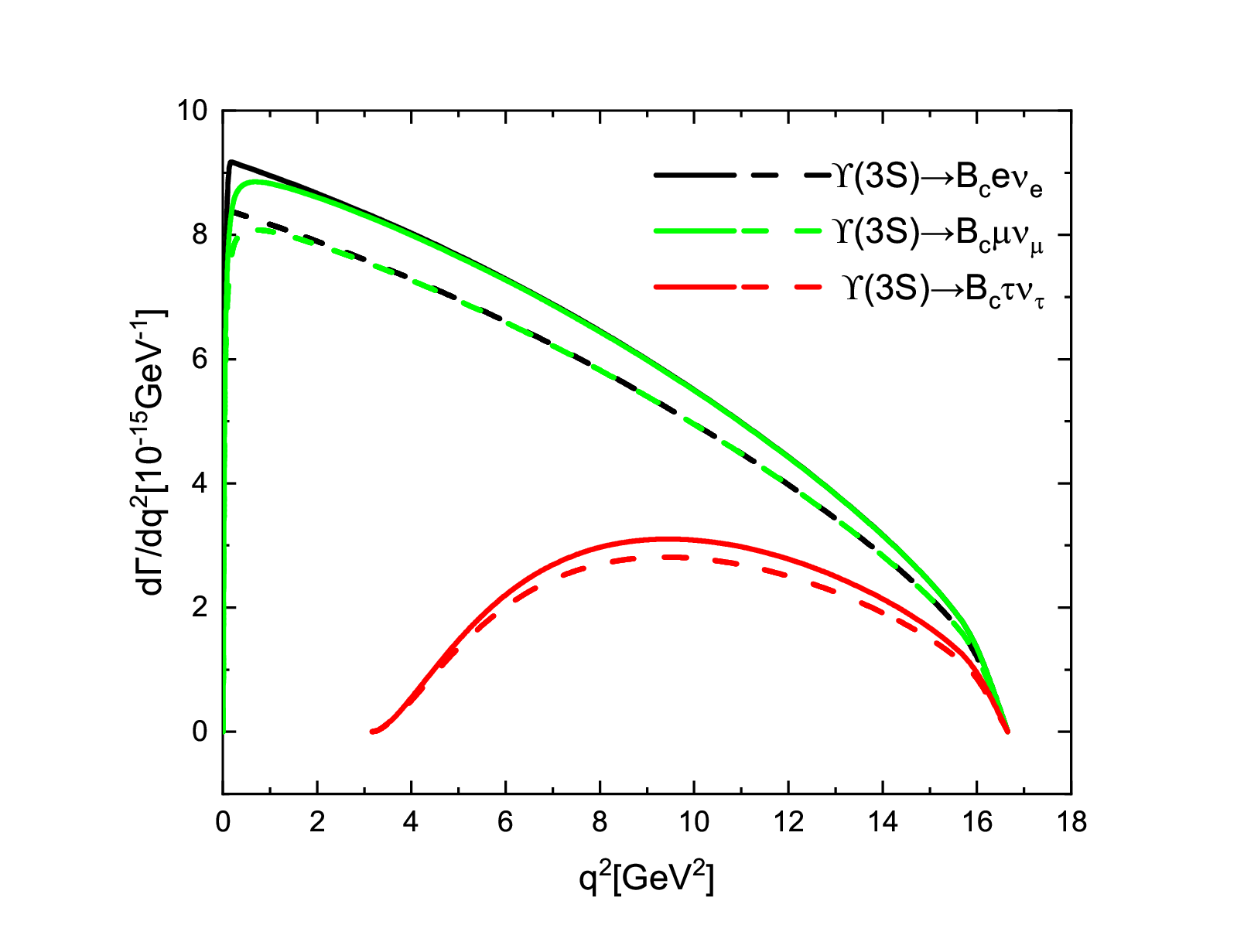}}
	\subfigure[]{\includegraphics[width=0.22\textwidth]{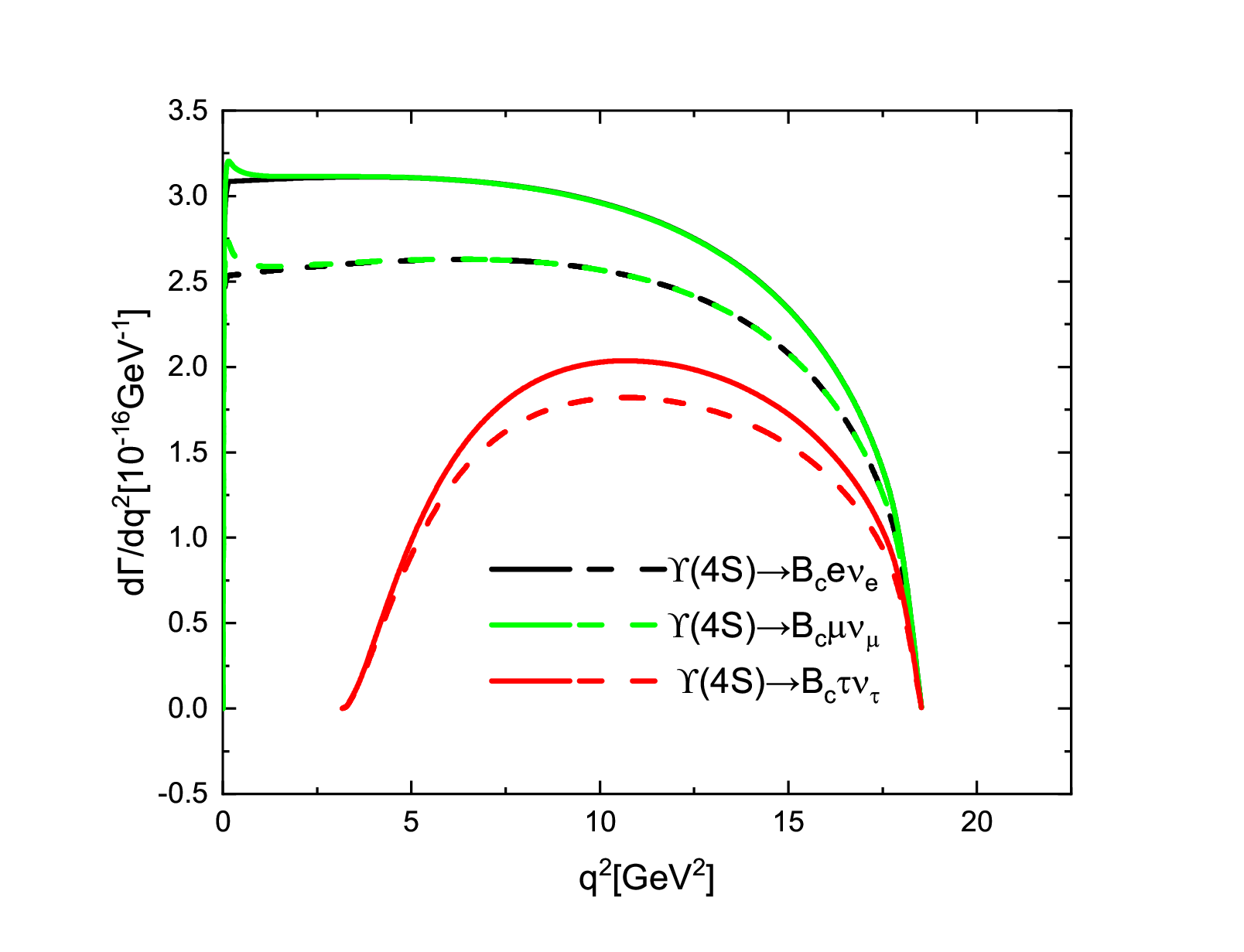}}\\
    \subfigure[]{\includegraphics[width=0.22\textwidth]{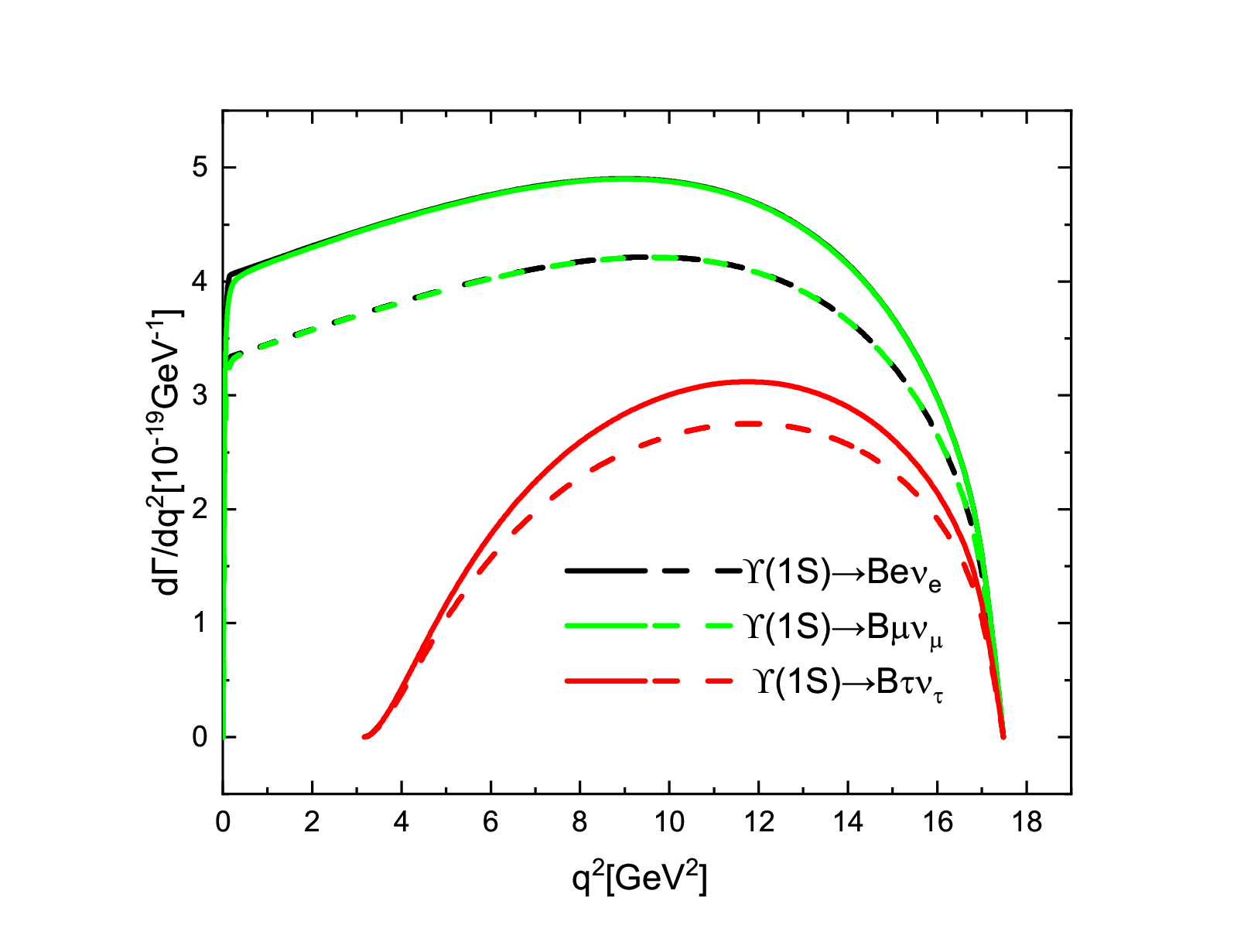}}
	\subfigure[]{\includegraphics[width=0.22\textwidth]{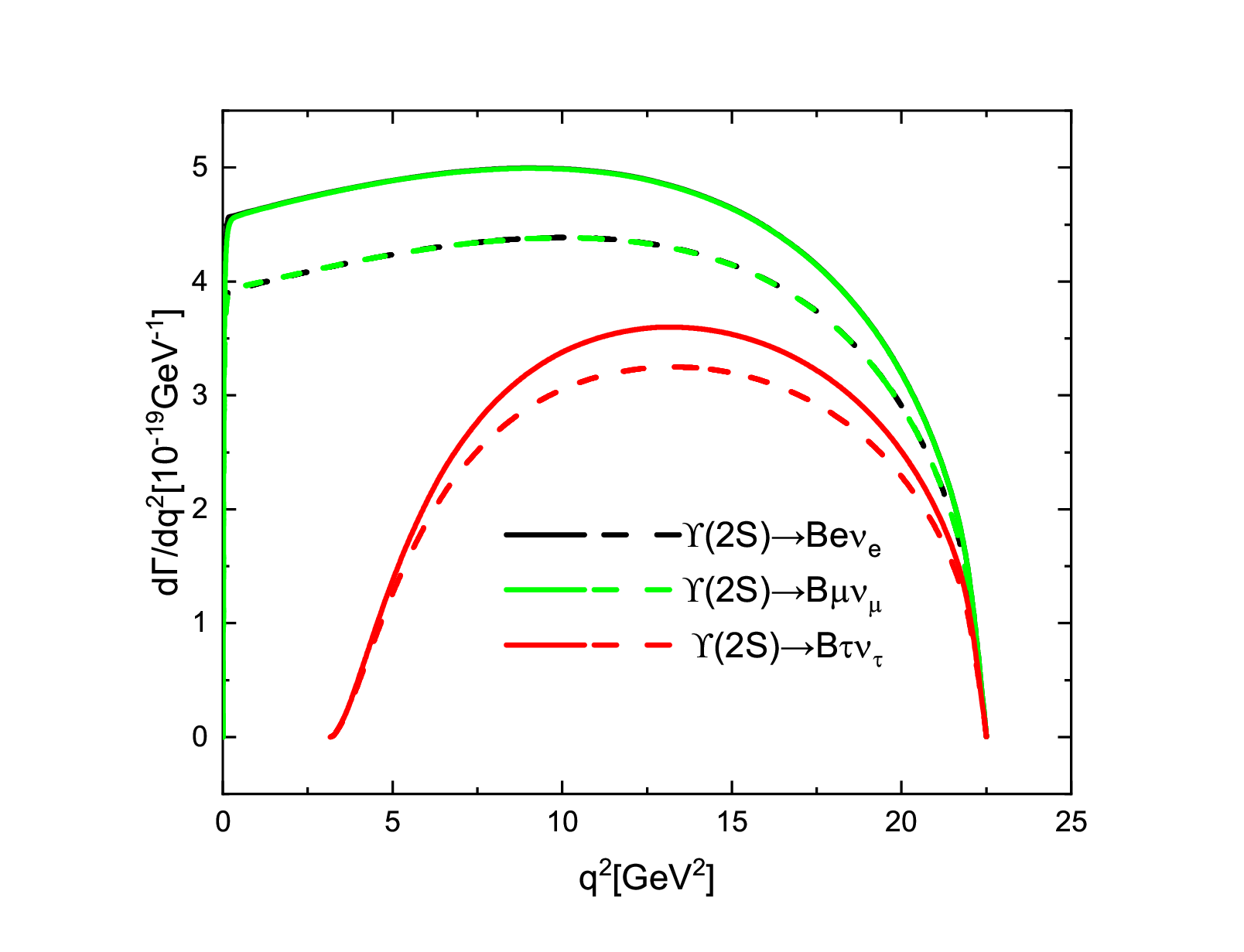}}
	\subfigure[]{\includegraphics[width=0.22\textwidth]{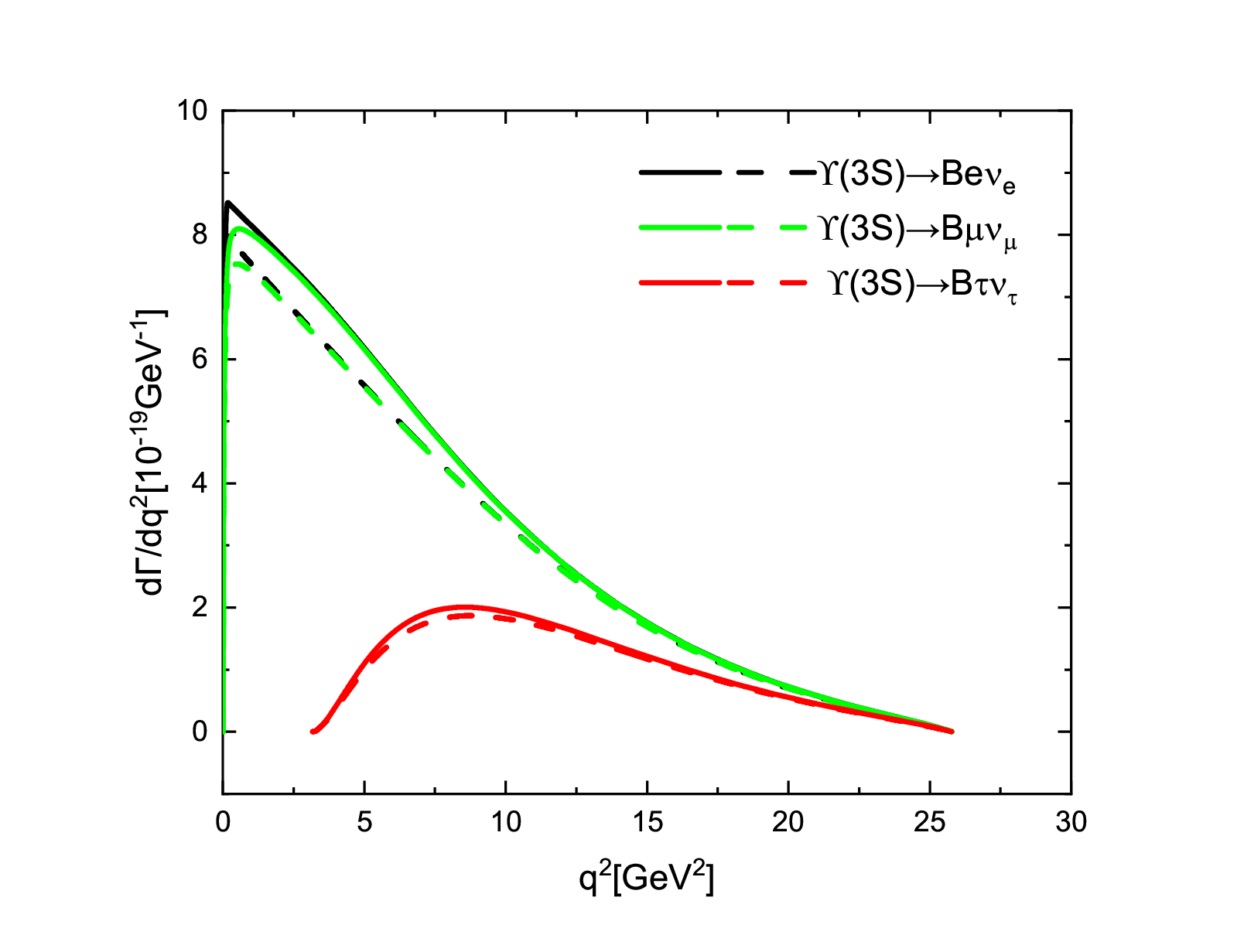}}
	\subfigure[]{\includegraphics[width=0.22\textwidth]{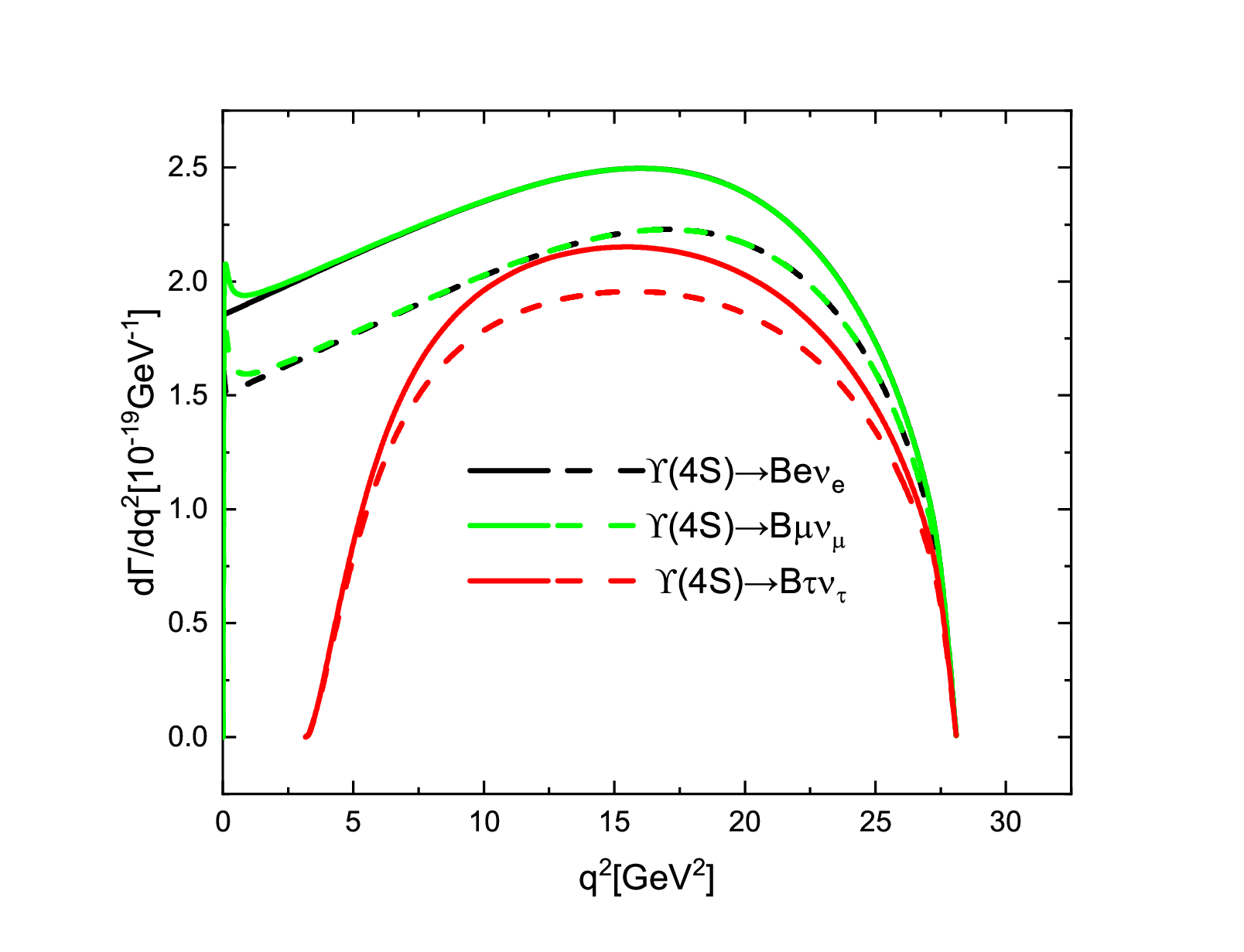}}
	\caption{The $q^2$ dependence of differential decay rates $d\Gamma^{(L)}/dq^2$, which are denoted as back, green and red solid (dashed) lines for the decays $\Upsilon(nS) \to B_{(c)} e {\nu}_{e}$, $\Upsilon(nS) \to B_{(c)} \mu {\nu}_{\mu}$ and $\Upsilon(nS) \to B_{(c)} \tau {\nu}_{\tau}$, respectively.}\label{fig:T2}
\end{figure}

\begin{table}[H]
	\caption{The longitudinal polarization fractions $f_{L}$ in Region 1, 2 and total $q^2$ region for the decays $\Upsilon(nS)\to B_{(c)}\ell\nu_{\ell}$.}
	\begin{center}
		\scalebox{0.9}{
			\begin{tabular}{|c|c|c|c||c|c|c|c|}
				\hline\hline
				Observables&Region 1&Region 2&Total&Observables&Region 1&Region 2&Total\\
				\hline\hline
				$f_{L}(\Upsilon(1S)\to B_{c}e\nu_{e})$&$0.831$&$0.833$&$0.832$&$f_{L}(\Upsilon(1S)\to B e\nu_{e})$&$0.840$&$0.874$&$0.856$\\
				\hline
				$f_{L}(\Upsilon(1S)\to B_{c} \mu\nu_{\mu})$&$0.833$&$0.833$&$0.833$&$f_{L}(\Upsilon(1S)\to B  \mu\nu_{\mu})$&$0.841$&$0.874$&$0.857$\\
				\hline
				$f_{L}(\Upsilon(1S)\to B_{c} \tau\nu_{\tau})$&$0.860$&$0.844$&$0.850$&$f_{L}(\Upsilon(1S)\to B  \tau\nu_{\tau})$&$0.882$&$0.886$&$0.884$\\
				\hline\hline
				$f_{L}(\Upsilon(2S)\to B_{c} e\nu_{e})$&$0.824$&$0.856$&$0.837$&$f_{L}(\Upsilon(2S)\to B e\nu_{e})$&$0.869$&$0.897$&$0.882$\\
				\hline
				$f_{L}(\Upsilon(2S)\to B_{c} \mu\nu_{\mu})$&$0.826$&$0.856$&$0.838$&$f_{L}(\Upsilon(2S)\to B  \mu\nu_{\mu})$&$0.870$&$0.897$&$0.882$\\
				\hline
				$f_{L}(\Upsilon(2S)\to B_{c} \tau\nu_{\tau})$&$0.883$&$0.873$&$0.877$&$f_{L}(\Upsilon(2S)\to B \tau\nu_{\tau})$&$0.906$&$0.907$&$0.907$\\
				\hline\hline
				$f_{L}(\Upsilon(3S)\to B_{c} e\nu_{e})$&$0.909$&$0.899$&$0.906$&$f_{L}(\Upsilon(3S)\to B e\nu_{e})$&$0.912$&$0.939$&$0.916$\\
				\hline
				$f_{L}(\Upsilon(3S)\to B_{c} \mu\nu_{\mu})$&$0.909$&$0.899$&$0.906$&$f_{L}(\Upsilon(3S)\to B  \mu\nu_{\mu})$&$0.913$&$0.965$&$0.921$\\
				\hline
				$f_{L}(\Upsilon(3S)\to B_{c} \tau\nu_{\tau})$&$0.913$&$0.965$&$0.921$&$f_{L}(\Upsilon(3S)\to B \tau\nu_{\tau})$&$0.930$&$0.967$&$0.941$\\
				\hline\hline
				$f_{L}(\Upsilon(4S)\to B_{c} e\nu_{e})$&$0.838$&$0.879$&$0.856$&$f_{L}(\Upsilon(4S)\to B e\nu_{e})$&$0.848$&$0.906$&$0.876$\\
				\hline
				$f_{L}(\Upsilon(4S)\to B_{c} \mu\nu_{\mu})$&$0.845$&$0.879$&$0.857$&$f_{L}(\Upsilon(4S)\to B \mu\nu_{\mu})$&$0.848$&$0.906$&$0.876$\\
				\hline
				$f_{L}(\Upsilon(4S)\to B_{c} \tau\nu_{\tau})$&$0.900$&$0.894$&$0.897$&$f_{L}(\Upsilon(4S)\to B \tau\nu_{\tau})$&$0.912$&$0.919$&$0.915$\\
				\hline\hline
			\end{tabular}\label{FL}}
	\end{center}
\end{table}
In Figure \ref{fig:T2}, it also shows that the longitudinal polarizations are dominant in these considered decays.

\subsection{Non-leptonic decays}
The decays rates of the nonleptonic weak decays $\Upsilon(nS)\to B_{(c)}M$ can be written as:
\be
\mathcal{B} r\left( \Upsilon(nS) \to B_{(c)} M\right)=\frac{p_{\mathrm{cm}}}{12 \pi m_{ \Upsilon}^{2} \Gamma_{ \Upsilon}}|\mathcal{A}(\Upsilon(nS) \to B_{(c)} M)|^{2},
\en
where $p_{cm}$ represents the three-momentum of the final meson $B_{(c)}$ or $M$ in the rest frame of $\Upsilon(nS)$.

\begin{table}[H]
	\caption{The branching ratios of the decays $\Upsilon(1S)\to B_{c}M$ with $M=\pi(\rho),K^{(*)},D^{(*)},D^{(*)}_s$. }
	\begin{center}
		\scalebox{0.75}{
			\begin{tabular}{|c |c |c |c|c |c |c |c |c |}
				\hline
				&This work&FH\cite{Sharma:1998gc}&BSW\cite{Dhir:2009rb}&QCDF \cite{Sun:2015faa}&PQCD\cite{Yang:2015iia,Sun:2016ndh,Sun:2016meb}&QCDF\cite{Sun:2015nya}&PQCD\cite{Sun:2015zoa}&QCDF\cite{Chang:2016enr}\\
				\hline
				$10^{-11}\times\mathcal{B} r(\Upsilon(1S)\to B_{c}^+\pi^-)$&$1.72^{+0.03+0.01+0.04}_{-0.03-0.01-0.04}$&$3.3$&$1.6$&$5.03$&$7.40$&$3.39$&$7.04$&$-$\\
				$10^{-12}\times\mathcal{B} r(\Upsilon(1S)\to B_{c}^+K^-)$&$1.30^{+0.02+0.01+0.03}_{-0.02-0.01-0.03}$&$2.4$&$1.3$&$3.73$&$5.67$&$2.51$&$5.41$&$-$\\
				$10^{-11}\times\mathcal{B} r(\Upsilon(1S)\to B_{c}^+\rho^-)$&$4.74^{+0.08+0.04+0.11}_{-0.08-0.04-0.11}$&$8.8$&$6.5$&$15.3$&$13.25$&$10.4$&$-$&$10.9$\\
				$10^{-12}\times\mathcal{B} r(\Upsilon(1S)\to B_{c}^+K^{\ast-})$&$2.80^{+0.05+0.02+0.07}_{-0.05-0.03-0.07}$&$5.0$&$3.5$&$8.75$&$7.97$&$5.26$&$-$&$6.4$\\
				$10^{-11}\times\mathcal{B} r(\Upsilon(1S)\to B_{c}^+D_{s}^-)$&$2.11^{+0.04+0.02+0.05}_{-0.04-0.03-0.05}$&$7.6$&$4.7$&$-$&$54.2$&$-$&$-$&$-$\\
				$10^{-13}\times\mathcal{B} r(\Upsilon(1S)\to B_{c}^+D^-)$&$7.79^{+0.16+0.09+0.18}_{-0.16-0.10-0.18}$&$31$&$15$&$-$&$196$&$-$&$-$&$-$\\
				$10^{-11}\times\mathcal{B} r(\Upsilon(1S)\to B_{c}^+D_{s}^{\ast-})$&$7.81^{+0.13+0.12+0.26}_{-0.14-0.14-0.24}$&$25.7$&$17.9$&$-$&$-$&$-$&$-$&$-$\\
				$10^{-12}\times\mathcal{B} r(\Upsilon(1S)\to B_{c}^+D^{\ast-})$&$3.36^{+0.06+0.05+0.11}_{-0.06-0.06-0.11}$&$11$&$7.3$&$-$&$-$&$-$&$-$&$-$\\
				\hline
			\end{tabular}\label{tab1}}
	\end{center}
\end{table}

Our numerical results are presented in Tables {\ref{tab1}}-{\ref{tab3}}, where the uncertainties come from the decay constants of the initial and final-state mesons and the lifetimes of $\Upsilon(nS)$, respectively. Between the decays $\Upsilon(nS)\to B_{c}\pi$ and $\Upsilon(nS)\to B_{c}K$, their branching ratios have a general rank-size relationship, that is $\mathcal{B} r(\Upsilon(nS)\to B_{c}\pi)>\mathcal{B} r(\Upsilon(nS)\to B_{c}K)$, which arises from the hierarchical structure of the CKM factors, namely $|V_{ud}|>|V_{us}|$. A similar situation arises when  $\pi (K)$ is replaced by $\rho (K^*)$ in these decays. 
More specifically, the branching ratios of all the decays $\Upsilon(nS)\to B_{c}\pi(\rho)$ are on the order of $10^{-11}$, those of almost all the decays
$\Upsilon(nS)\to B_{c}K(K^*)$ are on the order of $10^{-12}$. 
Another hierarchy of the CKM factors, $|V_{cs}|>|V_{cd}|$, leads to the rank-size relationship $\mathcal{B} r(\Upsilon(nS)\to B_{c}D^{(*)}_s)>\mathcal{B} r(\Upsilon(nS)\to B_{c}D^{(*)})$. In fact, the CKM matrix factors related to the decays $\Upsilon(nS)\to B_{c}\pi(\rho)$ are almost equal to those for the channels $\Upsilon(nS)\to B_{c}D^{(*)}_s$, so these two kinds of decays have similar branching ratios, whose difference originates from the discrepancies in the decay constants of the emitted mesons and phase space factors.  A similar relationship exists between the decays$\Upsilon(nS)\to B_{c}K(K^*)$ and $\Upsilon(nS)\to B_{c}D(D^*)$. Normally, there exists the following relation for these considered decays: $\mathcal{B} r(\Upsilon(3S)\to B_{c}M)>\mathcal{B} r(\Upsilon(2S)\to B_{c}M)> \mathcal{B}r(\Upsilon(1S)\to B_{c}M)$ for the same final meson M, due to the fact that $m_{\Upsilon(3S)} > m_{\Upsilon(2S)}> m_{\Upsilon(1S)}$ and $\Gamma_{\Upsilon(3S)}<\Gamma_{\Upsilon(2S)}<\Gamma_{\Upsilon(1S)}$. ‌It is noteworthy that‌  $\mathcal{B} r(\Upsilon(3S)\to B_{c}M)$ are significantly larger than $\mathcal{B} r(\Upsilon(2S)\to B_{c}M)$ 
by a factor of $3\sim5$. In a word,
the decays $\Upsilon(1S, 2S, 3S) \to B_c D^{(*)}_s$ and $\Upsilon(1S, 2S, 3S) \to B_c \pi(\rho)$ exhibit relatively large branching ratios within the range of $10^{-11}\sim 10^{-10}$. The $\Upsilon(ns)$ production cross sections in p-pb collisions are from a few to twenty $\mu b$ at LHCb . About $10^{12}\sim10^{13} \Upsilon(nS)$ data samples can be available per $ab^{-1}$ data, corresponding to dozens to about two hundred of $\Upsilon(1S, 2S, 3S) \to B_c D^{(*)}_s(\pi,\rho)$ decay events. If the lower identification efficiency (only about $1\%$) for the final-state meson $B_c$ is taken into account, which is identified through the decay $B_c\to J/\Psi \mu\nu_\mu$ or $J/\Psi\pi$ with branching ratios at the order of $10^{-4}\sim10^{-3}$, measuring these decays remains highly challenging even in future high-luminosity experiments. 
We also list the results from other theoretical approaches for comparison, such as the factorization hypothesis (FH), the BSW model, the QCD factorization (QCDF) and the PQCD.
One can find that our calculations for the decay channels $\Upsilon(1S) \to B_{c} M$ are comparable with those given in the BSW model, while most of our predictions are  smaller than the FH \cite{Sharma:1998gc}, the PQCD \cite{Yang:2015iia,Sun:2016ndh,Sun:2015zoa,Sun:2016meb} and the QCDF \cite{Sun:2015faa,Sun:2015nya,Chang:2016enr} results. However, the results for some decays, such as $\Upsilon(2S) \to B_{c} \rho(K^*)$ and $ \Upsilon(3S) \to B_{c}\pi(K)$, are consistent with those from the PQCD approach within errors. 
Although both the decays $\Upsilon(nS) \to B_{c}D_s^{(*)}, B_{c}D^{(*)}$ and $\Upsilon(nS) 
\to B_{c}\pi(\rho), B_cK^{(*)}$ are color-favored and CKM favored, their 
Feynman diagrams are very different: the latter only have emission topologies, and the former include annihilation topologies except for emission ones. 
Within the PQCD approach, the annihilation diagrams are expected to make 
considerable contributions to the branching ratios for some decay channels.  
So it is natural to understand that there exists significant difference between our predictions and the PQCD results for the decays $\Upsilon(nS) \to B_{c}D_s^{(*)}, B_{c}D^{(*)}$ as shown in Tables \ref{tab1} and \ref{tab2}. It is interesting for future experimental studies to test whether the annihilation diagrams play such a significant role. 

 The branching ratios of the decays $\Upsilon(4S) \to B_{c}M$ are very tiny and on the order of $10^{-14}\sim10^{-15}$ as shown in Table \ref{taby4} in Appendix D, due to the large decay width of the $\Upsilon(4S)$, which is about $10^{3}$ times as large as that of $\Upsilon(3S)$. Compared to the decays $\Upsilon(1S,2S,3S) \to B_cM$, the processes $\Upsilon(1S,2S,3S) \to BM$ are strongly suppressed by the CKM matrix elements, so their branching ratios are also very tiny and less than $10^{-14}$. It is naturally understandable that the branching ratios of the decays $\Upsilon(4S)\to BM$ are the smallest in those of corresponding decays $\Upsilon(nS)\to B_{(c)}M$. It is noted that the discrepancies between the branching ratios of the decays $\Upsilon(4S)\to BM$ and $\Upsilon(1S,2S,3S)\to BM$ are not as significant as those between $\Upsilon(4S)\to B_cM$ and $\Upsilon(1S,2S,3S)\to B_cM$, because the former is
 reduced by the relevant form factors.
\begin{table}[H]
	\caption{The branching ratios of the decays $\Upsilon(2S, 3S)\to B_{c}M$ with $M=\pi(\rho),K^{(*)},D^{(*)},D^{(*)}_s$. }
	\begin{center}
		\scalebox{1.05}{
			\begin{tabular}{|c |c |c |c |c  |}
				\hline
				&This work&PQCD\cite{Yang:2015iia,Sun:2016ndh,Sun:2016meb}&QCDF\cite{Sun:2015nya}&QCDF\cite{Chang:2016enr}\\
				\hline
				$10^{-11}\times\mathcal{B} r(\Upsilon(2S)\to B_{c}^+\pi^-)$&$2.26^{+0.09+0.01+0.20}_{-0.11-0.01-0.17}$&$6.29$&$8.27$&$-$\\
				$10^{-12}\times\mathcal{B} r(\Upsilon(2S)\to B_{c}^+K^-)$&$1.73^{+0.07+0.01+0.16}_{-0.08-0.01-0.13}$&$4.85$&$6.18$&$-$\\
				$10^{-11}\times\mathcal{B} r(\Upsilon(2S)\to B_{c}^+\rho^-)$&$5.79^{+0.22+0.04+0.52}_{-0.26-0.04-0.44}$&$8.88$&$24.7$&$25.9$\\
				$10^{-12}\times\mathcal{B} r(\Upsilon(2S)\to B_{c}^+K^{\ast-})$&$3.36^{+0.12+0.03+0.30}_{-0.15-0.02-0.26}$&$5.28$&$12.28$&$15.0$\\
				$10^{-11}\times\mathcal{B} r(\Upsilon(2S)\to B_{c}^+D_{s}^-)$&$3.74^{+0.20+0.04+0.33}_{-0.22-0.03-0.28}$&$42.8$&$-$&$-$\\
				$10^{-12}\times\mathcal{B} r(\Upsilon(2S)\to B_{c}^+D^-)$&$1.33^{+0.07+0.01+0.12}_{-0.08-0.01-0.10}$&$13.8$&$-$&$-$\\
				$10^{-11}\times\mathcal{B} r(\Upsilon(2S)\to B_{c}^+D_{s}^{\ast-})$&$8.80^{+0.31+0.12+0.79}_{-0.36-0.11-0.67}$&$-$&$-$&$-$\\
				$10^{-12}\times\mathcal{B} r(\Upsilon(2S)\to B_{c}^+D^{\ast-})$&$3.75^{+0.13+0.04+0.34}_{-0.15-0.05-0.28}$&$-$&$-$&$-$\\
				\hline
				$10^{-11}\times\mathcal{B} r(\Upsilon(3S)\to B_{c}^+\pi^-)$&$7.65^{+1.39+0.00+0.77}_{-1.22-0.00-0.64}$&$6.57$&$12.40$&$-$\\
				$10^{-12}\times\mathcal{B} r(\Upsilon(3S)\to B_{c}^+K^-)$&$5.81^{+1.05+0.00+0.58}_{-0.93-0.00-0.49}$&$5.09$&$9.30$&$-$\\
				$10^{-10}\times\mathcal{B} r(\Upsilon(3S)\to B_{c}^+\rho^-)$&$2.37^{+0.39+0.01+0.24}_{-0.35-0.01-0.20}$&$0.846$&$3.71$&$3.91$\\
				$10^{-11}\times\mathcal{B} r(\Upsilon(3S)\to B_{c}^+K^{\ast-})$&$1.44^{+0.23+0.01+0.14}_{-0.21-0.01-0.12}$&$0.498$&$1.909$&$2.26$\\
				$10^{-10}\times\mathcal{B} r(\Upsilon(3S)\to B_{c}^+D_{s}^-)$&$1.08^{+0.20+0.02+0.11}_{-0.18-0.01-0.09}$&$4.61$&$-$&$-$\\
				$10^{-12}\times\mathcal{B} r(\Upsilon(3S)\to B_{c}^+D^-)$&$3.88^{+0.72+0.05+0.39}_{-0.63-0.04-0.32}$&$15.8$&$-$&$-$\\
				$10^{-10}\times\mathcal{B} r(\Upsilon(3S)\to B_{c}^+D_{s}^{\ast-})$&$4.53^{+0.65+0.10+0.45}_{-0.59-0.09-0.38}$&$-$&$-$&$-$\\
				$10^{-11}\times\mathcal{B} r(\Upsilon(3S)\to B_{c}^+D^{\ast-})$&$1.95^{+0.28+0.04+0.19}_{-0.25-0.04-0.16}$&$-$&$-$&$-$\\
				\hline
			\end{tabular}
			\label{tab2}
			}
	\end{center}
\end{table}

\section{Summary}\label{sum}
In this work, we used the CLFQM approach to investigate the semileptonic and nonleptonic rare weak decays of $\Upsilon(nS)$ with $n=1,2,3,4$, which can provide an important reference for future related experimental research. Here, the form factors of the transitions $\Upsilon(nS)\to B_c$ and $\Upsilon(nS)\to B$ were calculated, which play a crucial role in the considered decays $\Upsilon(nS)\to B_{(c)}\ell\nu_\ell$ and $\Upsilon(nS)\to B_{(c)}M$. One can find that (I) The branching ratios of the decays $\Upsilon(3S)\to B_c\ell\nu_\ell$ are the largest among those of the decays $\Upsilon(nS)\to B_c\ell\nu_\ell$ and reach up to $10^{-9}$, due to the narrow width of $\Upsilon(3S)$ and the large form factor $V ^{\Upsilon(3S)\to B_c}$. The branching ratios of the decays $\Upsilon(1S)\to B_c\ell\nu_\ell$ and $\Upsilon(2S)\to B_c\ell\nu_\ell$ are close to each other and lie in the range $10^{-11}\sim10^{-10}$. (II) The forward-backward asymmetries $A_{FB}$ of the decays $\Upsilon(1S,2S,4S)\to B \ell\nu_\ell$ are about 1.6 or 2 times of the corresponding $A_{FB}(\Upsilon(1S,2S,4S)\to B_c \ell\nu_\ell)$ in magnitude, while there is an anomaly in the semileptonic decays
$\Upsilon(3S)\to B_{(c)} \ell\nu_\ell$, where the sizes of the
$A_{FB}(\Upsilon(3S)\to B \ell\nu_\ell)$ are smaller than those of the corresponding $A_{FB}(\Upsilon(3S)\to B_c \ell\nu_\ell)$. All the considered semileptonic decays $\Upsilon(nS)\to B_{(c)} \ell\nu_\ell$ are dominated by the longitudinal polarization and the polarization fractions are larger than $80\%$.
(III) Most of the branching ratios of the decays $\Upsilon(1S,2S,3S)\to B_cM$ are on the order of $10^{-11}$ or $10^{-12}$. Similar to the case of semileptonic decays, the largest branching ratios appear in the $\Upsilon(3S)$ decays, for example, 
$\Upsilon(3S)\to B^+_c\rho^-$ and $\Upsilon(3S)\to B^+_cD^{(*)-}_s$, whose branching ratios can amount to $10^{-10}$. (IV) Either the $\Upsilon(4S)$ decays or the channels involving $B$ meson considered here have tiny branching ratios, which are less than $10^{-13}$, due to the suppression from the wide width of $\Upsilon(4S)$ or/and the small CKM matrix element $V_{ub}$. 

\section*{Acknowledgment}
This work is partly supported by the National Natural Science Foundation of China under
Grant No. 11347030 and the Natural Science Foundation of Henan Province under grant
no. 232300420116, 252300421302.
\appendix
\section{Expressions of the trace $S_{\mu\nu}^{\Upsilon  B}$ }
The expression of the trace $S_{\mu\nu}^{\Upsilon  B}$ obtained by using the Lorentz contraction is written as 
\be
S_{\mu \nu}^{\Upsilon B}&=&\left(S_{V}^{\Upsilon B}-S_{A}^{\Upsilon B}\right)_{\mu \nu}\non
&=&\operatorname{Tr}\left[\left(\gamma_{\nu}-\frac{1}{W_{V}^{\prime \prime}}\left(p_{1}^{\prime \prime}-p_{2}\right)_{\nu}\right)\left(p_{1}^{\prime \prime}
+m_{1}^{\prime \prime}\right)\left(\gamma_{\mu}-\gamma_{\mu} \gamma_{5}\right)\left(\not p_{1}^{\prime}+m_{1}^{\prime}\right) \gamma_{5}\left(-\not p_{2}
+m_{2}\right)\right] \non
&=&-2 i \epsilon_{\mu \nu \alpha \beta}\left\{p_{1}^{\prime \alpha} P^{\beta}\left(m_{1}^{\prime \prime}-m_{1}^{\prime}\right)
+p_{1}^{\prime \alpha} q^{\beta}\left(m_{1}^{\prime \prime}+m_{1}^{\prime}-2 m_{2}\right)+q^{\alpha} P^{\beta} m_{1}^{\prime}\right\} \non
&&+\frac{1}{W_{V}^{\prime \prime}}\left(4 p_{1 \nu}^{\prime}-3 q_{\nu}-P_{\nu}\right) i \epsilon_{\mu \alpha \beta \rho} p_{1}^{\prime \alpha} q^{\beta} P^{\rho}\non &&
+2 g_{\mu \nu}\left\{m_{2}\left(q^{2}-N_{1}^{\prime}-N_{1}^{\prime \prime}-m_{1}^{\prime 2}-m_{1}^{\prime \prime 2}\right)
-m_{1}^{\prime}\left(M^{\prime \prime 2}-N_{1}^{\prime \prime}-N_{2}-m_{1}^{\prime \prime 2}-m_{2}^{2}\right)\right.\non
&&\left.-m_{1}^{\prime \prime}\left(M^{\prime 2}-N_{1}^{\prime}-N_{2}-m_{1}^{\prime 2}-m_{2}^{2}\right)
-2 m_{1}^{\prime} m_{1}^{\prime \prime} m_{2}\right\} \non &&
+8 p_{1 \mu}^{\prime} p_{1 \nu}^{\prime}\left(m_{2}-m_{1}^{\prime}\right)-2\left(P_{\mu} q_{\nu}
+q_{\mu} P_{\nu}+2 q_{\mu} q_{\nu}\right) m_{1}^{\prime}+2 p_{1 \mu}^{\prime} P_{\nu}\left(m_{1}^{\prime}-m_{1}^{\prime \prime}\right)\non &&
+2 p_{1 \mu}^{\prime} q_{\nu}\left(3 m_{1}^{\prime}-m_{1}^{\prime \prime}-2 m_{2}\right)
+2 P_{\mu} p_{1 \nu}^{\prime}\left(m_{1}^{\prime}+m_{1}^{\prime \prime}\right)+2 q_{\mu} p_{1 \nu}^{\prime}\left(3 m_{1}^{\prime}+m_{1}^{\prime \prime}-2 m_{2}\right)\non &&
+\frac{1}{2 W_{V}^{\prime \prime}}\left(4 p_{1 \nu}^{\prime}-3 q_{\nu}-P_{\nu}\right)\left\{2 p_{1 \mu}^{\prime}\left[M^{\prime 2}
+M^{\prime \prime 2}-q^{2}-2 N_{2}+2\left(m_{1}^{\prime}-m_{2}\right)\left(m_{1}^{\prime \prime}+m_{2}\right)\right]\right.\non&&
+q_{\mu}\left[q^{2}-2 M^{\prime 2}+N_{1}^{\prime}-N_{1}^{\prime \prime}+2 N_{2}-\left(m_{1}^{\prime}+m_{1}^{\prime \prime}\right)^{2}+2\left(m_{1}^{\prime}-m_{2}\right)^{2}\right]\non&&
\left.+P_{\mu}\left[q^{2}-N_{1}^{\prime}-N_{1}^{\prime \prime}-\left(m_{1}^{\prime}+m_{1}^{\prime \prime}\right)^{2}\right]\right\} .
\label{sptov}\en
\section{Some specific rules under the $p^-$ intergration}
When preforming the integraion, we need to include the zero-mode contribution. It amounts to performing the $p^-$ integration in a proper way in the CLFQM. Specificlly we
use the following rules given in Refs. \cite{jaus,Y. Cheng}
\be
\hat{p}_{1 \mu}^{\prime} &\doteq &   P_{\mu}
A_{1}^{(1)}+q_{\mu} A_{2}^{(1)},\\
\hat{p}_{1 \mu}^{\prime}
\hat{p}_{1 \nu}^{\prime}  &\doteq & g_{\mu \nu} A_{1}^{(2)} +P_{\mu}
P_{\nu} A_{2}^{(2)}+\left(P_{\mu} q_{\nu}+q_{\mu} P_{\nu}\right)
A_{3}^{(2)}+q_{\mu} q_{\nu} A_{4}^{(2)},\\
Z_{2}&=&\hat{N}_{1}^{\prime}+m_{1}^{\prime 2}-m_{2}^{2}+\left(1-2
x_{1}\right) M^{\prime 2} +\left(q^{2}+q \cdot P\right)
\frac{p_{\perp}^{\prime} \cdot q_{\perp}}{q^{2}},\\
\hat{p}_{1 \mu}^{\prime} \hat{N}_{2} & \rightarrow & q_{\mu}\left[A_{2}^{(1)} Z_{2}+\frac{q \cdot P}{q^{2}} A_{1}^{(2)}\right],\\
\hat{p}_{1 \mu}^{\prime} \hat{p}_{1 \nu}^{\prime} \hat{N}_{2} & \rightarrow &g_{\mu \nu} A_{1}^{(2)} Z_{2}+q_{\mu} q_{\nu}\left[A_{4}^{(2)} Z_{2}+2 \frac{q \cdot P}{q^{2}} A_{2}^{(1)} A_{1}^{(2)}\right],\\
A_{1}^{(1)}&=&\frac{x_{1}}{2}, \quad A_{2}^{(1)}=
A_{1}^{(1)}-\frac{p_{\perp}^{\prime} \cdot q_{\perp}}{q^{2}},\quad A_{3}^{(2)}=A_{1}^{(1)} A_{2}^{(1)},\\
A_{4}^{(2)}&=&\left(A_{2}^{(1)}\right)^{2}-\frac{1}{q^{2}}A_{1}^{(2)},\quad A_{1}^{(2)}=-p_{\perp}^{\prime 2}-\frac{\left(p_{\perp}^{\prime}
	\cdot q_{\perp}\right)^{2}}{q^{2}}, \quad A_{2}^{(2)}=\left(A_{1}^{(1)}\right)^{2}.  \en
\section{Eexpressions of the $\Upsilon(nS) \rightarrow B_{(c)}$ transition form factors}
The analytical expressions of the $\Upsilon(nS) \to B_{(c)}$ transition form factors in the CLFQM are given as following,
\begin{footnotesize}
\begin{eqnarray}
V(q^{2})&=&\frac{N_{c}(M^{'}+M^{''})}{16 \pi^{3}} \int d x_{2} d^{2} p_{\perp}^{\prime} \frac{2 h_{\Upsilon}^{\prime}
 h_{B_{(c)}}^{\prime \prime}}{x_{2} \hat{N}_{1}^{\prime} \hat{N}_{1}^{\prime \prime}}\left\{x_{2} m_{1}^{\prime}
 +x_{1} m_{2}+\left(m_{1}^{\prime}-m_{1}^{\prime \prime}\right) \frac{p_{\perp}^{\prime} \cdot q_{\perp}}{q^{2}}\right.\non &&\left.
 +\frac{2}{w_{V}^{\prime \prime}}\left[p_{\perp}^{\prime 2}+\frac{\left(p_{\perp}^{\prime} \cdot q_{\perp}\right)^{2}}{q^{2}}\right]\right\},\\
A_1(q^{2})&=& -\frac{1}{M^{'}+M^{''}}\frac{N_{c}}{16 \pi^{3}} \int d x_{2} d^{2} p_{\perp}^{\prime} \frac{h_{\Upsilon}^{\prime} h_{B_{(c)}}^{\prime \prime}}{x_{2}
\hat{N}_{1}^{\prime}
\hat{N}_{1}^{\prime \prime}}\left\{2 x_{1}\left(m_{2}-m_{1}^{\prime}\right)\left(M_{0}^{\prime 2}+M_{0}^{\prime \prime 2}\right)
-4 x_{1} m_{1}^{\prime \prime} M_{0}^{\prime 2}\right.\non
&&\left.+2 x_{2} m_{1}^{\prime} q \cdot P+2 m_{2} q^{2}-2 x_{1} m_{2}\left(M^{\prime 2}+M^{\prime \prime 2}\right)+2\left(m_{1}^{\prime}-m_{2}\right)\left(m_{1}^{\prime}
+m_{1}^{\prime \prime}\right)^{2}+8\left(m_{1}^{\prime}-m_{2}\right) \right.\non &&
\left. \times\left[p_{\perp}^{\prime 2}+\frac{\left(p_{\perp}^{\prime}
\cdot q_{\perp}\right)^{2}}{q^{2}}\right]+2\left(m_{1}^{\prime}+m_{1}^{\prime \prime}\right)\left(q^{2}+q \cdot P\right) \frac{p_{\perp}^{\prime} \cdot q_{\perp}}{q^{2}}
-4 \frac{q^{2} p_{\perp}^{\prime 2}+\left(p_{\perp}^{\prime} \cdot q_{\perp}\right)^{2}}{q^{2} w_{V}^{\prime \prime}}
\right.\non && \left.\times\left[2 x_{1}\left(M^{\prime 2}+M_{0}^{\prime 2}\right)-q^{2}-q \cdot P-2\left(q^{2}+q \cdot P\right) \frac{p_{\perp}^{\prime} \cdot q_{\perp}}{q^{2}}-2\left(m_{1}^{\prime}-m_{1}^{\prime \prime}\right)\left(m_{1}^{\prime}-m_{2}\right)\right]\right\},\;\;\;\;\;\\
A_2(q^{2})&=& \frac{N_{c}(M^{'}+M^{''})}{16 \pi^{3}} \int d x_{2} d^{2} p_{\perp}^{\prime} \frac{2 h_{\Upsilon}^{\prime} h_{B_{(c)}}^{\prime \prime}}{x_{2} \hat{N}_{1}^{\prime}
\hat{N}_{1}^{\prime \prime}}\left\{\left(x_{1}-x_{2}\right)\left(x_{2} m_{1}^{\prime}+x_{1} m_{2}\right)-\frac{p_{\perp}^{\prime} \cdot q_{\perp}}{q^{2}}\left[2 x_{1} m_{2}
+m_{1}^{\prime \prime} \right.\right.\non &&
\left.\left.+\left(x_{2}-x_{1}\right) m_{1}^{\prime}\right]-2 \frac{x_{2} q^{2}+p_{\perp}^{\prime} \cdot q_{\perp}}{x_{2} q^{2} w_{V}^{\prime \prime}}\left[p_{\perp}^{\prime} \cdot p_{\perp}^{\prime \prime}
+\left(x_{1} m_{2}+x_{2} m_{1}^{\prime}\right)\left(x_{1} m_{2}-x_{2} m_{1}^{\prime \prime}\right)\right]\right\},\\
A_0(q^{2})&=& \frac{M^{'}+M^{''}}{2M^{''}}A_1(q^{2})-\frac{M^{'}-M^{''}}{2M^{''}}A_2(q^{2})-\frac{q^2}{2M^{''}}\frac{N_{c}}{16 \pi^{3}} \int d x_{2} d^{2} p_{\perp}^{\prime} \frac{h_{\Upsilon}^{\prime} h_{B_{(c)}}^{\prime \prime}}{x_{2} \hat{N}_{1}^{\prime}
\hat{N}_{1}^{\prime \prime}}\left\{2\left(2 x_{1}-3\right)\right.\non &&\left.\times\left(x_{2} m_{1}^{\prime}+x_{1} m_{2}\right)-8\left(m_{1}^{\prime}-m_{2}\right)
\times\left[\frac{p_{\perp}^{\prime 2}}{q^{2}}
+2 \frac{\left(p_{\perp}^{\prime} \cdot q_{\perp}\right)^{2}}{q^{4}}\right]-\left[\left(14-12 x_{1}\right) m_{1}^{\prime}\right.\right. \non &&\left.\left.-2 m_{1}^{\prime \prime}-\left(8-12 x_{1}\right) m_{2}\right] \frac{p_{\perp}^{\prime} \cdot q_{\perp}}{q^{2}}
+\frac{4}{w_{V}^{\prime \prime}}\left(\left[M^{\prime 2}+M^{\prime \prime 2}-q^{2}+2\left(m_{1}^{\prime}-m_{2}\right)\left(m_{1}^{\prime \prime}
+m_{2}\right)\right]\right.\right.\non &&\left.\left.\times\left(A_{3}^{(2)}+A_{4}^{(2)}-A_{2}^{(1)}\right)
+Z_{2}\left(3 A_{2}^{(1)}-2 A_{4}^{(2)}-1\right)+\frac{1}{2}\left[x_{1}\left(q^{2}+q \cdot P\right)
-2 M^{\prime 2}-2 p_{\perp}^{\prime} \cdot q_{\perp}\right.\right.\right.\non &&\left.\left.\left.-2 m_{1}^{\prime}\left(m_{1}^{\prime \prime}+m_{2}\right)
-2 m_{2}\left(m_{1}^{\prime}-m_{2}\right)\right]\left(A_{1}^{(1)}+A_{2}^{(1)}-1\right) q \cdot P\left[\frac{p_{\perp}^{\prime 2}}{q^{2}}
+\frac{\left(p_{\perp}^{\prime} \cdot q_{\perp}\right)^{2}}{q^{4}}\right]\right.\right.\non &&\left.\left.\times\left(4 A_{2}^{(1)}-3\right)\right)\right\}.\;\;\;
\end{eqnarray}
\end{footnotesize}
\section{Branching ratios of the decays $\Upsilon(nS) \to B_{(c)}M$ with $M=\pi(\rho),K^{(*)},D^{(*)},D^{(*)}_s$ }
\begin{table}[H]
	\caption{The branching ratios of the decays $\Upsilon(4S)\to B_{c}M$ with $M=\pi(\rho),K^{(*)},D^{(*)},D^{(*)}_s$.  }
	\begin{center}
		\scalebox{0.9}{
			\begin{tabular}{|c |c |c |c |}
				\hline
				&This work&&This work\\
				\hline
				$10^{-14}\times\mathcal{B} r(\Upsilon(4S)\to B_{c}^+\pi^-)$&$2.62^{+0.04+0.02+0.36}_{-0.06-0.02-0.28}$&$10^{-14}\times\mathcal{B} r(\Upsilon(4S)\to B_{c}^+D_{s}^-)$&$5.10^{+0.15+0.06+0.71}_{-0.19-0.05-0.55}$\\
				$10^{-15}\times\mathcal{B} r(\Upsilon(4S)\to B_{c}^+K^-)$&$2.02^{+0.03+0.02+0.28}_{-0.05-0.02-0.22}$&$10^{-15}\times\mathcal{B} r(\Upsilon(4S)\to B_{c}^+D^-$&$1.76^{+0.05+0.03+0.46}_{-0.08-0.03-0.36}$\\
				$10^{-14}\times\mathcal{B} r(\Upsilon(4S)\to B_{c}^+\rho^-)$&$6.59^{+0.10+0.06+0.92}_{-0.15-0.06-0.72}$&$10^{-14}\times\mathcal{B} r(\Upsilon(4S)\to B_{c}^+D_{s}^{\ast-})$&$9.79^{+0.18+0.11+1.36}_{-0.25-0.13-1.06}$\\
				$10^{-15}\times\mathcal{B} r(\Upsilon(4S)\to B_{c}^+K^{\ast-})$&$3.81^{+0.06+0.03+0.53}_{-0.09-0.04-0.41}$&$10^{-15}\times\mathcal{B} r(\Upsilon(4S)\to B_{c}^+D^{\ast-})$&$4.16^{+0.07+0.05+0.58}_{-0.10-0.05-0.45}$\\
				\hline
			\end{tabular}\label{taby4}}
	\end{center}
\end{table}

\begin{table}[H]
	\caption{The branching ratios of the decays $\Upsilon(1S,2S,3S,4S)\to B M$ with $M=\pi(\rho),K^{(*)},D^{(*)},D^{(*)}_s$.  }
	\begin{center}
		\scalebox{0.9}{
			\begin{tabular}{|c |c ||c|c |}
				\hline
				&This work&&This work\\
				\hline
				$10^{-15}\times\mathcal{B} r(\Upsilon(1S)\to B^+\pi^-)$&$6.70^{+0.26+3.82+0.16}_{-0.25-2.85-0.15}$&$10^{-14}\times\mathcal{B} r(\Upsilon(2S)\to B^+\pi^-)$&$1.46^{+0.17+0.73+0.13}_{-0.17-0.58-0.11}$\\
				$10^{-16}\times\mathcal{B} r(\Upsilon(1S)\to B^+K^-)$&$5.02^{+0.20+2.90+0.12}_{-0.19-2.15-0.11}$&$10^{-15}\times\mathcal{B} r(\Upsilon(2S)\to B^+K^-)$&$1.10^{+0.13+0.56+0.10}_{-0.13-0.44-0.08}$\\
				$10^{-14}\times\mathcal{B} r(\Upsilon(1S)\to B^+\rho^-)$&$1.67^{+0.06+1.01+0.04}_{-0.06-0.73-0.04}$&$10^{-14}\times\mathcal{B} r(\Upsilon(2S)\to B^+\rho^-)$&$3.50^{+0.42+1.81+0.31}_{-0.42-1.42-0.27}$\\
				$10^{-16}\times\mathcal{B} r(\Upsilon(1S)\to B^+K^{\ast-})$&$9.53^{+0.37+5.87+0.23}_{-0.36-4.22-0.22}$&$10^{-15}\times\mathcal{B} r(\Upsilon(2S)\to B^+K^{\ast-})$&$1.99^{+0.24+1.04+0.18}_{-0.24-0.81-0.15}$\\
				$10^{-15}\times\mathcal{B} r(\Upsilon(1S)\to B^+D_{s}^-)$&$7.59^{+0.35+5.35+0.18}_{-0.34-3.62-0.17}$&$10^{-14}\times\mathcal{B} r(\Upsilon(2S)\to B^+D_{s}^-)$&$1.90^{+0.29+1.21+0.17}_{-0.28-0.87-0.14}$\\
				$10^{-16}\times\mathcal{B} r(\Upsilon(1S)\to B^+D^-)$&$2.79^{+0.13+1.93+0.07}_{-0.12-1.32-0.06}$&$10^{-16}\times\mathcal{B} r(\Upsilon(2S)\to B^+D^-)$&$6.89^{+1.04+4.29+0.62}_{-0.99-3.12-0.52}$\\
				$10^{-14}\times\mathcal{B} r(\Upsilon(1S)\to B^+D_{s}^{\ast-})$&$1.71^{+0.07+1.34+0.04}_{-0.07-0.85-0.04}$&$10^{-14}\times\mathcal{B} r(\Upsilon(2S)\to B^+D_{s}^{\ast-})$&$3.57^{+0.52+2.33+0.32}_{-0.49-1.67-0.27}$\\
				$10^{-16}\times\mathcal{B} r(\Upsilon(1S)\to B^+D^{\ast-})$&$7.66^{+0.33+5.84+0.18}_{-0.32-3.78-0.17}$&$10^{-15}\times\mathcal{B} r(\Upsilon(2S)\to B^+ D^{\ast-})$&$1.59^{+0.22+1.01+0.14}_{-0.22-0.73-0.12}$\\
				\hline\hline
				$10^{-16}\times\mathcal{B} r(\Upsilon(3S)\to B^+\pi^-)$&$3.58^{+4.60+2.04+0.36}_{-2.46-3.18-0.30}$&$10^{-17}\times\mathcal{B} r(\Upsilon(4S)\to B^+\pi^-)$&$1.96^{+0.17+0.79+0.03}_{-0.18-0.67-0.02}$\\
				$10^{-17}\times\mathcal{B} r(\Upsilon(3S)\to B^+K^-)$&$1.99^{+2.97+1.42+0.20}_{-1.48-1.45-0.17}$&$10^{-18}\times\mathcal{B} r(\Upsilon(4S)\to B^+K^-)$&$1.46^{+0.10+0.60+0.02}_{-0.04-0.51-0.02}$\\
			    $10^{-15}\times\mathcal{B} r(\Upsilon(3S)\to B^+\rho^-)$&$3.36^{+1.34+0.99+0.34}_{-0.86-2.48-0.28}$&$10^{-17}\times\mathcal{B} r(\Upsilon(4S)\to B^+\rho^-)$&$4.74^{+0.43+1.97+0.66}_{-0.45-1.65-0.52}$\\
				$10^{-16}\times\mathcal{B} r(\Upsilon(3S)\to B^+K^{\ast-})$&$2.18^{+0.78+0.62+0.22}_{-0.51-0.16-0.18}$&$10^{-18}\times\mathcal{B} r(\Upsilon(4S)\to B^+K^{\ast-})$&$2.70^{+0.25+1.13+0.38}_{-0.26-0.95-0.29}$\\
				$10^{-16}\times\mathcal{B} r(\Upsilon(3S)\to B^+D_s^-)$&$7.80^{+2.61+1.24+0.78}_{-2.15-0.78-0.65}$&$10^{-17}\times\mathcal{B} r(\Upsilon(4S)\to B^+D_{s}^-)$&$3.16^{+0.39+1.61+0.44}_{-0.40-1.27-0.34}$\\
				$10^{-17}\times\mathcal{B} r(\Upsilon(3S)\to B^+D^-)$&$1.97^{+0.77+0.43+0.20}_{-0.85-1.82-0.16}$&$10^{-18}\times\mathcal{B} r(\Upsilon(4S)\to B^+D^-)$&$1.12^{+0.14+0.56+0.16}_{-0.14-0.44-0.18}$\\
				$10^{-15}\times\mathcal{B} r(\Upsilon(3S)\to B^+D_s^{\ast-})$&$2.92^{+0.48+0.93+0.29}_{-0.34-2.32-0.24}$&$10^{-17}\times\mathcal{B} r(\Upsilon(4S)\to B^+D_{s}^{\ast-})$&$5.24^{+0.62+2.74+0.73}_{-0.62-2.15-0.57}$\\
				$10^{-16}\times\mathcal{B} r(\Upsilon(3S)\to B^+D^{\ast-})$&$1.41^{+0.27+0.47+0.14}_{-0.19-0.33-0.12}$&$10^{-18}\times\mathcal{B} r(\Upsilon(4S)\to B^+D^{\ast-})$&$2.30^{+0.26+1.17+0.32}_{-0.27-0.93-0.25}$\\
				\hline
			\end{tabular}\label{tab3}}
	\end{center}
\end{table}

\end{document}